\documentclass[nojss]{jss}

\usepackage{BOONDOX-cal} 
\usepackage{upgreek} 
\usepackage{bm} 
\usepackage{amsmath} 
\usepackage{etoolbox} 
\usepackage{nicematrix} 
\usepackage[ruled,vlined,linesnumbered]{algorithm2e} 
\usepackage{enumitem} 
\DeclareMathOperator*{\argmax}{arg\,max} 
\usepackage{orcidlink} 
\usepackage[nameinlink]{cleveref} 
\usepackage{tcolorbox} 
\usepackage{amssymb} 
\usepackage{tikz} 
\usepackage[bb=boondox]{mathalfa} 

\author{Jan H. R. Dressler~
        \orcidlink{0009-0005-7272-3173}\\
        Clausthal University of Technology
   \And Peter Kurz~
        \orcidlink{0000-0002-3697-5948}\\
        bms marketing research + strategy
   \AND Winfried J. Steiner~
        \orcidlink{0009-0009-6756-4576}\\
        Clausthal University of Technology}
\Plainauthor{Jan H. R. Dressler, 
             Peter Kurz, 
             Winfried J. Steiner}

\title{Discrete Choice and Competitive Reactions: End-to-End Simulation 
       with the \proglang{R} Package \pkg{cash}}
\Plaintitle{Discrete Choice and Competitive Reactions: End-to-End Simulation 
            with the R Package cash}
\Shorttitle{Discrete Choice and Competitive Reactions: E2E Simulation in 
            \proglang{R}}
            
\Abstract{
  Although discrete choice (choice-based conjoint) analysis has become a widely
  used technique for the elicitation of consumer preferences and hence a 
  foundation for product design, to the best of our knowledge, there exists 
  neither free and open-source nor commercial software that covers the 
  game-theoretic simulation of competitive reactions among firms based on 
  discrete choice models to improve decision making beyond traditional product 
  (line) optimization. The \proglang{R} package \pkg{cash} (\textbf{c}onjoint 
  $+$ N\textbf{ash}) does not only provide functions to fill this gap but 
  comprises an entire simulation pipeline including the upstream processes of 
  discrete choice analysis itself. \pkg{cash} ranges from preference generation, 
  choice design, error and response simulation, through Bayesian model 
  estimation and evaluation, to Nash equilibrium computation. Doing so, it 
  partly draws from established \proglang{R} packages concerned with discrete 
  choice analysis. While the structure of \pkg{cash} generally aims towards 
  end-to-end simulation as well as simulation of competitive dynamics based on 
  real data, all its key elements mentioned above may be of use independently of 
  each other.
}

\Plainkeywords{choice-based conjoint, discrete choice, mixed logit, 
               hierarchical Bayes, product line design, competitive reactions, 
               Nash equilibrium, R}

\Address{
  Jan H. R. Dressler, Winfried J. Steiner\\
  Department of Marketing\\
  Institute of Management, Economics and Law\\
  Clausthal University of Technology\\
  Julius-Albert-Strasse 2\\
  38678 Clausthal-Zellerfeld, Germany\\
  E-mail: \email{jhrd13@tu-clausthal.de}, 
          \email{winfried.steiner@tu-clausthal.de}\\
  \\
  Peter Kurz\\
  bms marketing research + strategy\\
  Landsberger Strasse 487\\
  81241 Munich, Germany\\
  E-mail: \email{p.kurz@bms-net.de}\\
}

\begin{document}

\section{Introduction} \label{sec:intro}
As much as product design is a creative task, most of the time it is also an 
optimization problem in terms of market share, revenue or profit. To maximize 
the latter objective, which encompasses the former two as special cases, it is 
necessary to determine the respective products' unit contribution margins as 
well as their expected (relative or absolute) demand. Since the pioneering works 
on conjoint and discrete choice analysis by \cite{1971_Green_Rao}, later Nobel 
laureate \cite{1974_McFadden} and \cite{1983_Louviere_Woodworth}, conjoint 
choice experiments have matured into a default for the elicitation of consumer 
preferences \citep[see, e.g.,][to get an impression of the many advances and 
applications]{2000_Louviere_et_al,2007_Gustafsson_et_al,2009_Train,2014_Rao,
2015_Hensher_et_al,2021_Baier_Brusch,2024_Hess_Daly}, which allow us to forecast
a product's share of choice (i.e., the expected relative demand) within a 
certain competitive scenario. More specifically, in such experiments, 
respondents are repeatedly asked to choose their preferred option from varying 
sets of alternatives resembling real purchasing situations. The resulting 
discrete choices are then decomposed into the respondents' individual utilities 
for the presented feature levels of the product by training correspondingly 
defined random utility models on the data. With these individual feature level 
utilities, it is possible to predict the respondents' choice behavior in a 
competitive scenario of alternatives and therefore the shares of choice required 
for product (line) optimization.

Now, after (successfully) launching a new or modified product (line), maybe due 
to the help of the above mentioned product (line) optimization, competitors will
most likely react to not lose market share, revenue or profit. Anticipating such 
reactions has the potential of sustainably strengthening a firm's competitive 
position. While there exists a substantial body of research on product (line) 
optimization \citep[see, e.g.,][for overviews]{2008_Belloni_et_al,
2024_Baier_Voekler}, this does not hold true for the simulation of competitive 
reactions bringing together discrete choice models and non-cooperative game 
theory \citep[see][for an overview]{2025_Dressler_et_al}, especially the 
\cite{1951_Nash} and \cite{1934_vonStackelberg} game-theoretic strategies. As 
far as we are aware, there is neither a free and open-source nor a commercial 
software at hand in this regard. Although the existing studies are inevitably 
built on code, no one seems to have developed and published an official package. 
Furthermore, only a fraction of them concentrated on the firms' strategic 
product policy, i.e., the more complex long-run competition on both product 
price and design (i.e., non-price features), none of which integrated fully 
Bayesian choice models into their simulation, let alone under different types of 
choice behavior (probabilistic versus deterministic) 
\citep[see][]{2025_Dressler_et_al}.

For this reason, we introduce the \proglang{R} \citep{2026_RCoreTeam} package 
\pkg{cash} (\textbf{c}onjoint $+$ N\textbf{ash}), which is available from the 
Comprehensive R Archive Network (CRAN) at 
\url{https://CRAN.R-project.org/package=cash} and can be installed and loaded
via 
\begin{Schunk}
\begin{Sinput}
R> install.packages("cash")
R> library(cash)
\end{Sinput}
\end{Schunk}
\pkg{cash} contributes functions for simulating Nash equilibria based on 
discrete choice models and additionally for the steps to reach the latter, 
thereby offering a tool for end-to-end simulation as well as simulation on the 
basis of external data, whether it be real or artificial. To combine these 
elements, it partly draws from established discrete choice \proglang{R} packages. 

Doing justice to the structure of \pkg{cash}, i.e., the viable end-to-end 
simulation and simultaneous option to leverage the functions individually (or 
in parts), the rest of this paper consists of ten sections, seven of them 
following the order of \pkg{cash}'s eight functions if employed as an ensemble 
(the last two functions share a section), each presenting the respective 
function's theoretical background and practical use as independently as possible 
without compromising their cohesiveness. As \pkg{cash} has recently been put to 
the test in a large-scale Monte Carlo study by us demonstrating its potential, 
the theoretical background of the functions is taken from the methodological
framework of the corresponding paper \citep{2025_Dressler_et_al}. In particular, 
the functions are concerned with preference generation (\Cref{sec:prefgen}), 
choice design (\Cref{sec:choicedes}), error and response simulation 
(\Cref{sec:responsesim}), hierarchical Bayesian estimation of mixed logit models 
(\Cref{sec:hbmxlest}), design assessment (\Cref{sec:designeval}), convergence 
and model assessment (\Cref{sec:modeleval}), and pre-computations and Nash 
competition (\Cref{sec:nashgame}). Before diving into the functions, 
\Cref{sec:prelim} gives some preliminary information. \Cref{sec:sharerevopt} 
discusses the adjustments required to consider competition on market share or 
revenue instead of profit. \Cref{sec:limdev} addresses \pkg{cash}'s limitations 
and closes with ideas for future development.

\section{Preliminaries} 
\label{sec:prelim}
In this preliminary section, we would like to briefly explain two overarching 
aspects of the package that were initially intended for internal use only but 
are now still present because they might be helpful to the user.

1. The functions (as well as their output files) are labelled from \code{B_} to 
\code{I_} ($+$ name) for reasons of clarity regarding their logical order. 
\code{A_} is reserved for the (potential) file calling the functions.

2. Instead of returning its output to be stored in a variable, each function 
outputs multiple .rds files (change each function's argument \code{temp} to 
\code{FALSE} to not save into the session's temporary but the current working 
directory) to be re-imported by subsequent functions. These output files will be 
described in the respective sections, but there exists a single one that all 
functions share, called \code{A-I_finaloutput.rds}. It is a \code{list} of 
\code{lists} which is propagated through the functions, ultimately containing 
eight \code{list} objects, one for each of the eight functions: \code{$fun1}, 
..., \code{$fun8}. The elements of each of those eight \code{list} 
objects are 
\setlist{nolistsep}
  \begin{itemize}[noitemsep]
    \item \code{$name}, showing the name of the executed function,
    \item \code{$param}, being a \code{list} of the arguments' values at 
          execution,
    \item \code{$rdsname}, comprising the names of the function's (other) .rds 
          output files,
    \item \code{$rdscon} (possibly multiple numbered ones), carrying the content 
          of these .rds output files, and
    \item \code{$additionalcon} (optional), with additional information.
  \end{itemize}
In other words, \code{A-I_finaloutput.rds} merges the functions' in- and 
outputs. As such a file takes up a lot of additional storage depending on the 
functions' inputs (and therefore outputs), many elements are deliberately set to 
\code{NULL} because they are less important, too large and/or already included. 
This way, \code{A-I_finaloutput.rds} still outlines everything that exists in
the \pkg{cash} environment to the current point in time, but the file size stays 
negligibly small. If the information covered by \code{A-I_finaloutput.rds} suits
the user's needs, some or all of the individual .rds files may be deleted to
save storage and neaten the directory. This can be quite handy if the functions 
are run hundreds or thousands of times as an ensemble 
\citep[see][]{2025_Dressler_et_al}. However, the first of the eight functions 
must always be executed once to initiate \code{A-I_finaloutput.rds} and avoid 
subsequent importing errors. Note that if a function is called twice in the same
directory, its .rds files including the respective \code{$fun} in 
\code{A-I_finaloutput.rds} are overwritten.

\section{Preference generation} 
\label{sec:prefgen}
\subsection{Theoretical background}
\label{sec:prefgentheo}
As in many fields of science, there are numerous unsolved questions in the 
domain of discrete choice analysis and its extensions which are impossible to 
answer without simulated data. In \cite{2025_Dressler_et_al}, for example, we 
tackled the issue to which degree state-of-the-art discrete choice models are
capable of uncovering the true Nash equilibria arising under the true consumer 
preferences. Obviously, true consumer preferences are not known unless they are 
simulated. To still obtain valid outcomes in such an endeavor, we think it is of 
great importance to generate synthetic utilities approximating estimates seen in 
empirical research and commercial applications. Derived from the latter, 
\cite{2010a_Wirth,2010b_Wirth} proposed an approach for sampling means and 
variances of a multivariate normal preference structure that has recently been 
employed by \cite{2019_Hein_et_al,2020_Hein_et_al,2022_Hein_et_al,
2024_Goeken_et_al,2025_Dressler_et_al} after re-examination of resulting 
densities based on extensive practical experience.

For the theoretical explanation, let $l\in L=\{1,...,\mathcal{l}\}$ be the 
feature index, $m\in M=\{1,...,\mathcal{m}\}$ be the level index, 
$o\in O=\{1,...,\mathcal{o}\}$ be the parameter index such that 
$\mathcal{o}=\mathcal{l}(\mathcal{m}-1)$, and $\lfloor{\cdot}\rceil$ be the 
rounding to the nearest integer function. The definition of $\mathcal{o}$ 
implies modelling of main effects only and having discrete features (including
price) with a constant number of levels (not a prerequisite). Since firms often 
only use certain price points, it seems reasonable to assume discreteness for 
the price feature as well. Following \cite{2010a_Wirth,2010b_Wirth}, we draw 
$\lfloor{0.1\mathcal{o}}\rceil$, $\mathcal{o}-2\lfloor{0.1\mathcal{o}}\rceil$,
$\lfloor{0.1\mathcal{o}}\rceil$ random samples from the continuous uniform 
densities $U(-5,-2)$, $U(-2,2)$, $U(2,5)$, respectively, and concatenate them to
a vector $\bar{\bm{\upbeta}}$ of preference means. Simultaneously, $\mathcal{o}$ 
realizations of a random variable $X$ are taken as preference variances, where
\begin{gather}
  \begin{gathered}
    \label{prefsimu}
      X=\min(Y+Z_1,Z_2),\\
      Y\sim\Gamma(\kappa,\theta),\\
      Z_1\sim U(\zeta_1,\xi_1),\\
      Z_2\sim U(\zeta_2,\xi_2).
  \end{gathered}
\end{gather}
The Gamma density $\Gamma$, with shape $\kappa$ and scale $\theta$, was fit to 
empirical data by \cite{2010a_Wirth,2010b_Wirth} and truncated using 
realizations of the uniformly distributed continuous random variables 
$Z_{(\cdot)}$. Two resulting tuples of distributional parameters allow for a 
less and a more heterogeneous structure of preference variances to emerge 
($0.7,1.5,0.08,0.4,9,11$ and $0.7,4.5,0.2,2,13,18$ for 
$\kappa,\theta,\zeta_1,\xi_1,\zeta_2,\xi_2$). With the realizations of $X$ 
stacked into a vector $\bm{\upsigma}^2$, each pair of components from 
$\bar{\bm{\upbeta}}$ and $\bm{\upsigma}^2$ is subsequently used to draw a vector 
$\bm{\upbeta}_o$ of $\mathcal{i}$ random samples from a normal density, where 
$i\in I=\{1,...,\mathcal{i}\}$ denotes the respondent index. This leads to the 
preferences being dispersed differently within each feature level, which is 
arguably more realistic than setting a constant variance.

We then choose to assign the vectors $\bm{\upbeta}_o \forall o$ to the feature
levels by combining them to an $\mathcal{i}\times\mathcal{o}$ preference matrix 
$\bm{\mathrm{B}}$ according to the order given by $\mathcal{o}$ random samples
without replacement from a discrete uniform density $D(1,\mathcal{o})$. If we
wish to define a feature $l$ with monotonously changing part-worths 
corresponding to a specific sorting of its levels (e.g., price), we actually do
the preceding steps for $\mathcal{m}$ instead of $\mathcal{m}-1$ levels of $l$, 
rearrange the $\mathcal{m}$ entries of $l$ in $\bm{\mathrm{B}}$ in order for
each $i$, shift them so that the first column of $l$ becomes a null vector, and
lastly remove the latter from $\bm{\mathrm{B}}$. This is to avoid any violation 
of such a monotonicity condition by the reference category when the coding 
scheme of the design matrix is expected to be dummy in the response simulation
and the re-estimation.

\begin{figure}[!ht]
  \centering 
  \includegraphics{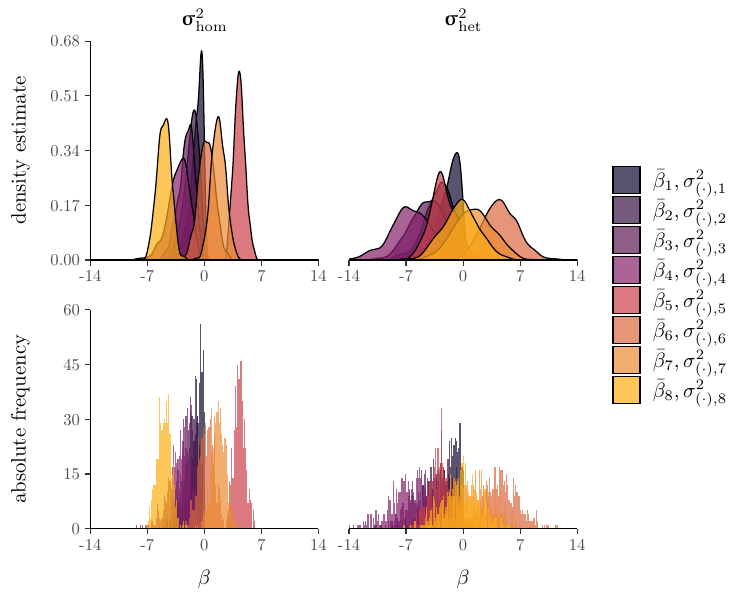} 
  \caption{Histograms and kernel density estimators of synthetic utilities}
  \label{fig:synutils}
\end{figure}

\Cref{fig:synutils} gives a typical example of homogeneous and heterogeneous 
synthetic utilities generated with the above method. Precisely, it displays 
histograms and kernel density estimators (using Gaussian kernels and Silverman's 
\citeyear{1986_Silverman} bandwidth) for $\mathcal{l}(\mathcal{m}-1)=2(5-1)=8$ 
parameters based on $\mathcal{i}=500$ individuals, showing plausible differences 
in heterogeneity and preference order without violating a monotonicity 
constraint exemplarily imposed on $l=1$ (violet densities). It can also be seen 
that the mixtures of normals resulting from this interference are not a matter 
of concern.

\subsection{Practical use}
\label{sec:prefgenprac}
\textit{\textbf{Function and input}}
\newline
The corresponding function from \pkg{cash} is \code{B_prefgen()}:
\begin{Code}
B_prefgen(resps, lvls, het = TRUE, randomizecols = TRUE,
  sortfirstattdesc = TRUE, saveplot = FALSE, seed = 1, temp = TRUE)
\end{Code}

\code{resps} defines the number of respondents and \code{lvls} the number of 
features and levels for which we would like to generate synthetic utilities. 
While \code{resps} takes an integer, \code{lvls} has to be given a vector of 
integers. The length of the vector is interpreted as the number of features and
the values as the number of levels per feature. The function automatically drops 
a level per feature, as the reference level does not have to be simulated, so it 
is necessary to specify the actual number of levels per feature. Except for 
\code{seed} (always addresses \code{set.seed()} in \pkg{cash}), the other 
arguments can be either \code{TRUE} or \code{FALSE}. If \code{het = TRUE}, the
utilities of each feature level are more heterogeneous across respondents. If 
\code{randomizecols = TRUE}, the generated column vectors containing the 
utilities for all respondents for each feature level are randomized (recall that 
the means for the feature levels are drawn in order from the continuous uniform 
densities $U(-5,-2)$, $U(-2,2)$, $U(2,5)$). If \code{sortfirstattdesc = TRUE}, 
the utilities of the first feature are sorted descendingly (taking the reference 
level (first level) utility of zero into account). We primarily implemented the 
latter to be able to impose a monotonicity constraint on the price feature. 
Since the first feature is expected to be price within \pkg{cash}, ordered from 
lowest to highest price, with \code{sortfirstattdesc = TRUE}, the highest 
utility corresponds to the lowest price and so forth. If \code{saveplot = TRUE}, 
a very basic \code{ggplot} \citep{2016_Wickham} of the final densities is saved.
\begin{tcolorbox}[width=0.95\textwidth,halign=justify,center,leftrule=2mm,
                  rounded corners,enlarge bottom finally by=5mm]
  \textbf{Important note}: The first feature is expected to be price within 
  \pkg{cash} (exceptions will be discussed in \Cref{sec:sharerevopt}). While 
  this may be inconsequential for the first functions (depending on the 
  settings), it is vital for later ones to work properly, so it is best to 
  consider it at this stage already. Also keep in mind that the first level will 
  be the reference level within each feature.
\end{tcolorbox}

\textit{\textbf{Output and example}}
\newline
Besides \code{A-I_finaloutput.rds} and the optional \code{ggplot}, 
\code{B_prefgen()} outputs the two .rds files
\setlist{nolistsep}
  \begin{itemize}[noitemsep]
    \item \code{B_simulatedindividualbetas.rds} and
    \item \code{B_levels.rds}.
  \end{itemize}

In the following example, the former is a $50\times 8$ (50 respondents $\times$ 
2 features with 5 levels each minus the reference level for each feature, first 
feature is price with a monotonicity constraint) heterogeneous preference matrix 
and the latter is the vector \code{c(5, 5)} given to the argument \code{lvls}.
\begin{tcolorbox}[width=0.95\textwidth,halign=justify,center,leftrule=2mm,
                  rounded corners]
  \textbf{Important note}: W.l.o.g., some of the settings in the examples of 
  this paper, like the low number of respondents and product features, are 
  chosen to ensure fast execution times and low memory usage of the entire 
  function ensemble on usual personal computers. More demanding settings can be 
  found in \cite{2025_Dressler_et_al}.
\end{tcolorbox}
\begin{Schunk}
\begin{Sinput}
R> B_prefgen(resps = 50, lvls = c(5, 5), temp = FALSE)
\end{Sinput}
\end{Schunk}
\begin{Schunk}
\begin{Sinput}
R> simulatedindividualbetas = readRDS("./B_simulatedindividualbetas.rds")
R> dim(simulatedindividualbetas)
\end{Sinput}
\begin{Soutput}
[1] 50  8
\end{Soutput}
\begin{Sinput}
R> simulatedindividualbetas[1:5, 1:5]
\end{Sinput}
\begin{Soutput}
       beta1      beta2      beta3     beta4      beta5
1 -2.1620762 -2.7612604 -3.5526438 -6.252955  2.5930359
2 -1.0771541 -1.1122947 -1.5977042 -2.635264 -2.1829506
3 -1.5198762 -3.6762063 -4.2548052 -4.441788  0.7529233
4 -0.5476048 -0.7509054 -0.9873948 -1.546882 -1.4259378
5 -2.3962619 -4.7498928 -5.6630598 -7.265818 -2.5207036
\end{Soutput}
\end{Schunk}

\section{Choice design} 
\label{sec:choicedes}
\subsection{Theoretical background}
\label{sec:choicedestheo}
With the ongoing objective of preserving generality, in our view, the generation 
of choice designs for the estimation of main effects based on synthetic 
utilities needs to strive for optimality from both the statistical and practical 
perspective as usual, although artificial respondents certainly are quite 
forgiving with respect to some of the otherwise highly important decisions to be 
made at this stage. Therefore, the selection of the numbers of levels, 
alternatives and choice sets should serve to facilitate meeting respondents' 
capabilities (hypothetically) as well as three well-known principles of 
efficient choice designs, namely orthogonality, level balance and minimal 
overlap \citep{1996_Huber_Zwerina,2010_Zwerina_et_al}. While staying in the 
commonly recommended ranges considering cognitive, behavioral and cost aspects 
\citep[see, e.g.,][]{1996_Johnson_Orme,1997_Pinnell_Englert,2001_Hensher_et_al,
2005_Caussade_et_al,2006_Hoogerbrugge_vanderWagt,2007_Haaijer_Wedel,
2012_Kurz_Binner,2013_Louviere_et_al,2019_Hein_et_al}, in 
\cite{2025_Dressler_et_al}, for example, we defined the number of levels to be 
constant across features, the number of alternatives to be equal to the number 
of levels and the number of choice sets to be a multiple of the latter (which, 
of course, is often not feasible and sometimes not even desirable in real-world 
settings). Symmetric designs (i.e., a constant number of levels across features) 
may also enable (hypothetical) control of a possible number of levels effect 
\citep{1981_Currim_et_al,1982_Wittink_et_al,1990_Wittink_et_al,
1990_Green_Srinivasan}.

After these pre-adjustments (which are the responsibility of the user), an 
established option is to apply the modified Fedorov algorithm 
\citep{1972_Fedorov,1980_Cook_Nachtsheim} to generate choice designs by 
maximizing D-efficiency, the most frequently employed statistical optimality 
criterion therefor \citep{2019_Street_Viney}. As it cannot be ruled out that the 
candidate set itself prohibits reaching optimal D-efficiency in case an 
orthogonal array is used, the candidate set of alternatives from which the 
algorithm swaps may be the full factorial if computational feasibility is given. 
Additionally, initiating from multiple random start designs is recommended 
\citep{2010_Kuhfeld}. Further, note that the Fisher information matrix depends 
on the parameters in logit models \citep{2019_Street_Viney} such that a prior 
assumption has to be made for maximizing D-efficiency. If the prior is set to 
zero, utility balance as the fourth principle of efficient choice designs is 
only trivially satisfied \citep{2010_Zwerina_et_al}. Specifying plausible 
non-zero (Bayesian) priors and blocking larger designs 
\citep[see, e.g.,][]{2009_Rose_Bliemer,2018_Walker_et_al} may be considered to 
improve efficiency. Lastly, and motivated solely by the practical perspective, 
the avoidance of set duplicates seems sensible.

Designs for predictive testing (hold-outs) may be generated using the same 
procedure, with the uniqueness constraint ideally being extended to the union of
training and test sets. Given the artificial nature of the respondents, there is 
no need for randomizing the sets.
\newpage

\subsection{Practical use}
\label{sec:choicedesprac}
\textit{\textbf{Function and input}}
\newline
The corresponding function from \pkg{cash} is \code{C_choicedes()}:
\begin{Code}
C_choicedes(lvls = "./B_levels.rds", setstraining, altstraining,
  setstest = setstraining, altstest = altstraining, 
  prior = "./B_simulatedindividualbetas.rds", zeroprior = TRUE,
  bayesian = FALSE, createdesign = TRUE, algorithm = c("modifiedfedorov", 
  "coordinateexchange"), trace = FALSE, startdesignstotest = 12,
  orthoarrayascandmodfed = FALSE, externaltrainingdesign, externaltestdesign,
  codingofexternal = c("dummy", "none"), efficiencyofexternaltraining = "",
  efficiencyofexternaltest = "", csvseparator = ";", seed = 1, temp = TRUE)
\end{Code}

First, we need to define the number of features and levels via the argument 
\code{lvls}, equivalent to \code{B_prefgen()}, but this time automatically 
through the corresponding .rds file coming from \code{B_prefgen()} (or an own 
one, of course) containing the vector. Second, it is necessary to specify the 
desired number of choice sets and alternatives per choice set via
\code{setstraining}, \code{altstraining} (for estimation), \code{setstest}, 
\code{altstest} (for predictive testing). Third, \code{prior} always has to be a 
Bayesian prior initially, i.e., an $\mathcal{i}\times\mathcal{o}$ preference 
matrix $\bm{\mathrm{B}}$ as outputted by \code{B_prefgen()}. If 
\code{zeroprior = TRUE}, the entries of this matrix are set to zero, and if 
\code{bayesian = FALSE}, its column means are taken as the (non-Bayesian) prior. 

For the actual generation of the choice designs (\code{createdesign = TRUE}), 
\code{C_choicedes()} borrows functions from the \proglang{R} package 
\pkg{idefix} \citep{2020_Traets_et_al} in reduced form. As in \pkg{idefix}, we 
can choose between \code{algorithm = "modifiedfedorov"} and (faster but possibly 
less efficient) \code{"coordinateexchange"} (\code{trace = TRUE} enables tracing 
for both algorithms) 
\citep[see][and the reference manual for details]{2020_Traets_et_al}. The number 
of random start designs to test can be given to \code{startdesignstotest}, and
by setting \code{orthoarrayascandmodfed = TRUE}, an orthogonal array generated 
with the \proglang{R} package \pkg{DoE.base} \citep{2018_Groemping} is employed 
as the candidate set instead of the full factorial in case of the modified 
Fedorov algorithm.

The user is explicitly invited to make use of \pkg{idefix}'s full potential 
(e.g., blocking larger designs) or other software \citep[e.g., the \proglang{SAS} 
\code{\%ChoicEff} macro (exclusion of set duplicates possible), see][]
{2010_Kuhfeld,2010_Zwerina_et_al} by importing externally generated designs. To
do so, they must come as .csv files with each row being an alternative (i.e., in 
long-format), row names as first column and column names as first row (the value
separator is controllable by \code{csvseparator} and the names themselves are 
irrelevant because they are redone internally). The arguments for the paths to 
the .csv files are \code{externaltrainingdesign} and \code{externaltestdesign}. 
If the external designs are already dummy-coded, \code{codingofexternal} must be
set to \code{"dummy"}, and if no coding scheme has been applied (i.e., each 
column corresponds to a feature and the entries are integers representing the 
levels), \code{codingofexternal} must be set to \code{"none"} to convert them to 
dummy-coding (first level becomes reference level).

As \code{C_choicedes()} needs the information for the above and other internal 
processes (e.g., the creation of respondent and choice IDs), the correct 
specification of \code{lvls}, \code{setstraining}, \code{altstraining}, 
\code{setstest}, \code{altstest} and \code{prior} is also mandatory if 
\code{createdesign = FALSE} and external designs are imported. Only for the sake
of completeness, efficiency measures of the external designs may be stored in 
the arguments \code{efficiencyof[...]}.

If not realized by the external designs (since it is also not included in 
\code{C_choicedes()} through \pkg{idefix}), the exclusion of set duplicates 
within and between designs may be checked with \code{F_designeval()} (see 
\Cref{sec:designeval}). There is a reason for \code{F_} not being named 
\code{D_}, which will be elucidated in the respective section, but note that 
\code{F_} is an exception in regard to the logical order from the practical
perspective and can be directly called after \code{C_} as one would expect.
\newline

\textit{\textbf{Output and example}}
\newline
\code{C_choicedes()} outputs the two .rds files
\setlist{nolistsep}
  \begin{itemize}[noitemsep]
    \item \code{C_trainingdesign.rds} and
    \item \code{C_testdesign.rds},
  \end{itemize}
each being a \code{list} with the design matrix stored in \code{$design} in 
long-format and the D(B)-error in \code{$error} (among other information). 

In the first of the two examples below, designs are generated with 
\code{C_choicedes()} for 2 features with 5 levels each (see \code{B_levels.rds} 
from \code{B_prefgen()} in \Cref{sec:prefgen}), 15 choice sets and 5 
alternatives per choice set.
\begin{Schunk}
\begin{Sinput}
R> C_choicedes(setstraining = 15, altstraining = 5, temp = FALSE)
\end{Sinput}
\end{Schunk}
\begin{Schunk}
\begin{Sinput}
R> trainingdesign = readRDS("./C_trainingdesign.rds")
R> trainingdesign$design[1:6, ]
\end{Sinput}
\begin{Soutput}
          Var12 Var13 Var14 Var15 Var22 Var23 Var24 Var25
set1.alt1     0     0     1     0     0     0     1     0
set1.alt2     0     0     0     0     0     1     0     0
set1.alt3     0     1     0     0     1     0     0     0
set1.alt4     1     0     0     0     0     0     0     1
set1.alt5     0     0     0     1     0     0     0     0
set2.alt1     0     0     1     0     0     0     1     0
\end{Soutput}
\end{Schunk}
For the second example, let us import the following external .csv designs,
ext1.csv and ext2.csv \citep[generated with the 
\proglang{SAS}\textsuperscript{\textregistered} \code{\%ChoicEff} macro,][]
{2010_Kuhfeld,2010_Zwerina_et_al}\footnote{\copyright\ 2024 SAS Institute Inc. 
SAS and all other SAS Institute Inc. product or service names are registered 
trademarks or trademarks of SAS Institute Inc., Cary, NC, USA.}, where ext1 (on 
the left, for training) has the same specifications as the first example and 
ext2 (on the right, for testing) only consists of 5 choice sets:
\begin{Code}
setalt;feat1;feat2 | setalt;feat1;feat2
11;5;2             | 11;1;1
12;1;3             | 12;4;2
13;3;5             | 13;5;3
14;4;4             | 14;2;5
15;2;1             | 15;3;4
21;2;5             | 21;1;5
...                | ...
\end{Code}
Here, the row and column names are meaningful, but that does not matter.
\begin{Schunk}
\begin{Sinput}
R> C_choicedes(setstraining = 15, altstraining = 5, setstest = 5, 
+    createdesign = FALSE, externaltrainingdesign = "./ext1.csv", 
+    externaltestdesign = "./ext2.csv", codingofexternal = "none", 
+    efficiencyofexternaltraining = "99.5536", 
+    efficiencyofexternaltest = "99.5515", temp = FALSE)
\end{Sinput}
\end{Schunk}
\begin{Schunk}
\begin{Sinput}
R> trainingdesign = readRDS("./C_trainingdesign.rds")
R> trainingdesign$design[1:6, ]
\end{Sinput}
\begin{Soutput}
          Var12 Var13 Var14 Var15 Var22 Var23 Var24 Var25
set1.alt1     0     0     0     1     1     0     0     0
set1.alt2     0     0     0     0     0     1     0     0
set1.alt3     0     1     0     0     0     0     0     1
set1.alt4     0     0     1     0     0     0     1     0
set1.alt5     1     0     0     0     0     0     0     0
set2.alt1     1     0     0     0     0     0     0     1
\end{Soutput}
\end{Schunk}

\section{Error and response simulation} 
\label{sec:responsesim}
\subsection{Theoretical background}
\label{sec:responsesimtheo}
Next, following random utility theory, noise must be added to the individual 
deterministic portion of utility of the artificial respondents across 
alternatives and choice sets prior to the simulation of choices. For the
generation of these random error terms, previous simulation studies fixed a 
scalar with which to multiply the standard variance of the chosen error density
\citep[see, e.g.,][]{2002a_Andrews_et_al,2003_Andrews_Currim,2010a_Wirth,
2010b_Wirth,2020_Hein_et_al} or set the variance of the latter to a percentage 
of the combined variance of the deterministic and the stochastic portion of 
utility \citep[see, e.g.,][]{1975_Srinivasan,1981_Wittink_Cattin,
1989_Wedel_Steenkamp,1996_Vriens_et_al,2002b_Andrews_et_al}. The second approach 
is far more flexible as it automatically adjusts for changes in the variability 
of the deterministic utility (e.g., when the number of features serves as an 
experimental factor). Nonetheless, explicitly defining a ratio of measures of 
dispersion still seems rather intangible because firstly, the inherent emphasis 
on larger values has to be taken into account when using variance or standard 
deviation. Secondly, expectation, skewness and kurtosis of both distributions 
(deterministic utility and error) are at least computationally neglected, and 
thirdly, the stochasticity in the subsequent assignment to the deterministic 
utilities is completely unrestricted.

In our view, it is preferable to directly control the proportion of error to 
deterministic utility, but to do so, the error has to be independent from the 
expectation of the deterministic utilities. Fortunately, in the case of 
zero-centering, the deterministic utilities (as well as the errors) themselves 
are deviations in their entirety, which makes the absence of translational 
invariance irrelevant. Otherwise, proportionate errors would actually be too 
large, while not zero-centering the errors might even make it impossible to 
sample proportionate and bidirectional ones. Therefore, zero-centering is a
prerequisite for both.

Based on these thoughts, we propose controlling post hoc for the median of 
proportions of error to deterministic utility, which we call the median relative 
Gumbel error (MRGE). The distributional choice is arbitrary and just driven by
the nature of the logit model used in \pkg{cash}. Let $\bm{\mathrm{x}}_{jk}$ be
a $1\times \mathcal{o}$ vector of dummy variables defining the alternative 
$j\in J=\{1,...,\mathcal{j}\}$ in choice set $k\in K=\{1,...,\mathcal{k}\}$
such that $\bm{\upbeta}_i\bm{\mathrm{x}}_{jk}^T$ is the deterministic utility of 
respondent $i$ for $j$ in $k$ (linear-additive part-worth model), and let 
$\varepsilon_{ijk}$ be a realization of the random error 
$E\sim \mathrm{Gumbel}(\lambda,s)$ with location $\lambda$ and scale $s$ 
\citep[to draw $\varepsilon_{ijk}$, we use the \proglang{R} package \pkg{evd},]
[]{2002_Stephenson}. As usual, an individual's total utility is assumed to be
$\bm{\upbeta}_i\bm{\mathrm{x}}_{jk}^T+\varepsilon_{ijk}$. Now, if
$\bm{\mathrm{v}}_{\mathrm{abs}}$ and $\bm{\mathrm{\varepsilon}}_{\mathrm{abs}}$
denote the vectors containing $|\bm{\upbeta}_i\bm{\mathrm{x}}_{jk}^T$| and
$|\varepsilon_{ijk}|$ $\forall i,j,k$, respectively, and $\oslash$ represents 
the Hadamard division, then the MRGE can be expressed as
\begin{gather}
  \begin{gathered}
    \label{MRGE}
      \mathrm{MRGE}=
        \mathrm{med}(\bm{\mathrm{\varepsilon}}_{\mathrm{abs}}
                     \oslash\bm{\mathrm{v}}_{\mathrm{abs}}).
  \end{gathered}
\end{gather}
The MRGE attempts to enhance interpretability by employing a more accessible 
fraction (e.g., let the error be 50\% of the deterministic utility on average) 
and to increase robustness through its post hoc inspection. The median is used 
to prevent the errors from being scaled down without the suppression of 
outliers.

Applying the tuning procedure below (\Cref{MRGEtuning}), the standard 
deviation $\sigma$ of the error density is iteratively adjusted until a required 
MRGE target value holds. Besides $\bm{\mathrm{v}}_{\mathrm{abs}}$, its mean
$\bar{v}_{\mathrm{abs}}$ and the MRGE target value, it takes a learning rate
$d$, a tolerance $t$ and a maximum number of iterations $r_\mathrm{{max}}$ as
inputs, which do not have to be precisely calibrated to guarantee functionality. 
\begin{center}
  \begin{minipage}{\linewidth}
  \IncMargin{1.5em}
    \begin{algorithm}[H]
      \SetAlCapHSkip{.7em}
      \footnotesize
        \SetAlCapFnt{\footnotesize}
        \SetAlCapNameFnt{\footnotesize}
        \SetAlgoNlRelativeSize{-2}
      \DontPrintSemicolon
      \SetKwBlock{Repeat}{repeat}{}
        $\bm{\mathrm{input}}$: 
          $\bm{\mathrm{v}}_{\mathrm{abs}}$, $\bar{v}_{\mathrm{abs}}$, 
          $\mathrm{MRGE_{target}}$, $d$, $t$, $r_\mathrm{{max}}$\;
        $h:=\mathrm{MRGE_{target}}$, $r:=1$, $\gamma:=$ Euler's constant\;
        \Repeat{
          $\sigma:=h\cdot\bar{v}_{\mathrm{abs}}$\;
          $s:=\sigma\cdot\frac{\sqrt{6}}{\pi}$\;
          $\lambda:=-s\gamma$\;
          draw $\bm{\mathrm{\varepsilon}}$ from $\mathrm{Gumbel}(\lambda,s)$\;
          $\bm{\mathrm{g}}$ $:=$ 
$\bm{\mathrm{\varepsilon}}_{\mathrm{abs}}\oslash\bm{\mathrm{v}}_{\mathrm{abs}}$\;
          remove undefined quotients from $\bm{\mathrm{g}}$\; 
          $\mathrm{MRGE_{actual}}:=\mathrm{med}(\bm{\mathrm{g}})$\;
          \uIf{$|\mathrm{MRGE_{actual}}-\mathrm{MRGE_{target}}|\le t$}{
            return $\bm{\mathrm{\varepsilon}}$\;
          }\Else{
            $h:=h+(\mathrm{MRGE_{target}}-\mathrm{MRGE_{actual}})\cdot d$\;
            \If{$h<0$}{
              $h:=z$, $Z\sim U(0,1)$\;
            }
          }
          $r:=r+1$\;
          \If{$r>r_{\mathrm{max}}$}{
            return adjust $d$, $t$, $r_{\mathrm{max}}$\;
          }
        }
      \caption{MRGE tuning procedure}\label{MRGEtuning}
    \end{algorithm}
  \end{minipage}
\end{center}

After having generated the errors, the choices are simulated based on the
resulting total utilities and the first choice rule,
\begin{gather}
  \begin{gathered}
    \label{fc}
      f_{ijk}=
        \begin{cases}
          1,\,\bm{\upbeta}_i\bm{\mathrm{x}}_{jk}^T+\varepsilon_{ijk}=
              \max_{j'}(\bm{\upbeta}_i\bm{\mathrm{x}}_{j'k}^T+
                \varepsilon_{ij'k}),\\
          0,\,\text{otherwise},
        \end{cases}
      !\exists j'\in J(j'=j).
  \end{gathered}
\end{gather}

\subsection{Practical use}
\label{sec:responsesimprac}
\textit{\textbf{Function and input}}
\newline
The corresponding function from \pkg{cash} is \code{D_responsesim()}:
\begin{Code}
D_responsesim(
  simulatedindividualbetas = "./B_simulatedindividualbetas.rds", 
  trainingdesign = "./C_trainingdesign.rds", 
  testdesign = "./C_testdesign.rds", MRGEtarget, learningrate = 0.5, 
  stoppingcriterion = 10^(-5), maxiterations = 10^4, transform = TRUE, 
  simulate = TRUE, seed = 1, temp = TRUE)
\end{Code}

As the goal is to let our artificial respondents answer our choice designs,  
\code{D_responsesim()}, by default, imports the synthetic utilities from 
\code{B_} as well as the choice designs from \code{C_} through the accordingly 
named first three arguments. Although this data does not necessarily have to 
originate from \code{B_} and \code{C_}, it must always be in .rds format and 
exhibit the identical structure. Fortunately, the option of end-to-end 
simulation allows the user to take a look at such .rds defaults by simply 
generating them with the preceding functions. 

While $\bm{\mathrm{v}}_{\mathrm{abs}}$ and $\bar{v}_{\mathrm{abs}}$ as inputs 
for the error simulation (see \Cref{MRGEtuning}) are automatically determined 
via the loaded preference matrix, MRGEtarget, $d$, $t$ and $r_\mathrm{{max}}$ 
must be defined explicitly. We preset well-functioning values for $d$ 
(\code{learningrate}), $t$ (\code{stoppingcriterion}) and $r_\mathrm{{max}}$ 
(\code{maxiterations}) employed in \cite{2025_Dressler_et_al} such that these 
arguments may be left unchanged, but MRGEtarget (\code{MRGEtarget}), i.e., the 
desired magnitude of the error, has to be set to a percentage of the 
deterministic utility (so to speak).

If \code{transform = TRUE}, \code{D_responsesim()} stores the training design 
and responses in an additional way such that they can be processed by the second 
and last external discrete choice \proglang{R} package we draw from in 
\pkg{cash}, namely \pkg{bayesm} \citep{2025_Rossi}, which is used for model 
estimation in \code{E_} (see \Cref{sec:hbmxlest}). With \code{simulate = FALSE}, 
it is possible to skip the error and response simulation to prepare external 
(experimental or differently simulated) data for model estimation. To do so, the 
imported design .rds files have to be identical to the ones generated by 
\code{C_} (\code{$design}, \code{$resps}, \code{$betas}, \code{$lvls}, 
\code{$sets...}, \code{$alts...}, \code{$respondentid...}, \code{$choiceid...},
rest can be \code{NA}), but \code{$design} must comprise the designs of all 
respondents with the responses in the last column of this matrix.
\newline

\textit{\textbf{Output and example}}
\newline
\code{D_responsesim()} therefore outputs up to three .rds files, which are
\setlist{nolistsep}
  \begin{itemize}[noitemsep]
    \item \code{D_trainingdesignwithsimuresponses.rds},
    \item \code{D_testdesignwithsimuresponses.rds} and
    \item \code{D_bayesmdatatraining.rds}.
  \end{itemize}
The first two are the \code{$design} elements from the \code{C_} outputs, 
replicated for each respondent and extended by two columns, the simulated total 
utilities and the first choices. The last file is similar to the first one but 
tailored to the \pkg{bayesm} requirements.

To give an example, the following call of \code{D_responsesim()} generates 
answers to the choice designs from \code{C_} through the use of the 
deterministic utilities from \code{B_} and an additional MRGE of 30\%.
\newpage
\begin{Schunk}
\begin{Sinput}
R> D_responsesim(MRGEtarget = 0.3, temp = FALSE)
\end{Sinput}
\end{Schunk}
\begin{Schunk}
\begin{Sinput}
R> trainingdesignwithsimuresponses =
+    readRDS("./D_trainingdesignwithsimuresponses.rds")
R> trainingdesignwithsimuresponses[1:7, -c(1:3)]
\end{Sinput}
\begin{Soutput}
               Var15 Var22 Var23 Var24 Var25      sim u sim fc
res1.set1.alt1     1     1     0     0     0 -3.3340463      0
res1.set1.alt2     0     0     1     0     0  3.8646304      1
res1.set1.alt3     0     0     0     0     1  0.6413848      0
res1.set1.alt4     0     0     0     1     0 -7.4063880      0
res1.set1.alt5     0     0     0     0     0  2.0602403      0
res1.set2.alt1     0     0     0     0     1 -3.8477016      0
res1.set2.alt2     0     0     1     0     0 -2.6594139      0
\end{Soutput}
\end{Schunk}
To be clear with regard to \code{simulate}, we would get the same
\code{D_bayesmdatatraining.rds} as with the previous settings if we now imported 
an .rds with the above matrix (without \code{sim u}) in \code{$design} and set 
\code{simulate = FALSE}.

\section{Hierarchical Bayesian estimation of mixed logit models} 
\label{sec:hbmxlest}
\subsection{Theoretical background}
\label{sec:hbmxlesttheo}
As stated in \Cref{sec:responsesim}, we let the stochastic portion of utility 
be i.i.d. Gumbel such that the individual's choice probability $\eta_{ijk}$ of 
choosing alternative $j$ in choice set $k$ given $\bm{\upbeta}_i$ is defined by 
the conditional logit model \citep{1974_McFadden}
\begin{gather}
  \begin{gathered}
    \label{mnl}
      \eta_{ijk}(\bm{\upbeta}_i)=
        \frac{\mathrm{exp}(\bm{\upbeta}_i\bm{\mathrm{x}}_{jk}^T)}
             {\sum_{j'}\mathrm{exp}(\bm{\upbeta}_i\bm{\mathrm{x}}_{j'k}^T)}.
  \end{gathered}
\end{gather}
For the coefficients $\bm{\upbeta}_i$, a multivariate normal prior $\phi$ with 
$\mathcal{o}\times1$ mean vector $\bar{\bm{\upbeta}}$ and 
$\mathcal{o}\times\mathcal{o}$ covariance matrix $\bm{\Sigma}$ is specified 
which describes the population's unimodal preference heterogeneity and leads to
the (unconditional) mixed logit probability \citep{1999_Brownstone_Train}
\begin{gather}
  \begin{gathered}
    \label{mlm}
      \mathcal{p}_{ijk}(\bar{\bm{\upbeta}},\bm{\Sigma})=
        \int
          \eta_{ijk}(\bm{\upbeta}_i)
          \phi(\bm{\upbeta}_i|\bar{\bm{\upbeta}},\bm{\Sigma})
        d\bm{\upbeta}_i.
  \end{gathered}
\end{gather}
Based on this well-established single component approach for the distribution of 
$\bm{\upbeta}_i$, the likelihood of $i$'s choice sequence $\bm{\mathrm{y}}_i$ 
then becomes
\begin{gather}
  \begin{gathered}
    \label{likelihood}
      \mathcal{L}(\bm{\mathrm{y}}_i|\bar{\bm{\upbeta}},\bm{\Sigma})=
        \int
          \mathcal{L}(\bm{\mathrm{y}}_i|\bm{\upbeta}_i)
          \phi(\bm{\upbeta}_i|\bar{\bm{\upbeta}},\bm{\Sigma})
        d\bm{\upbeta}_i=
        \int
          \bigg(\prod_{k,j}\eta_{ijk}(\bm{\upbeta}_i)^{f_{ijk}}\bigg)
          \phi(\bm{\upbeta}_i|\bar{\bm{\upbeta}},\bm{\Sigma})
        d\bm{\upbeta}_i.
  \end{gathered}
\end{gather}

For the Bayesian estimation of the hyperparameters $\bar{\bm{\upbeta}}$ and
$\bm{\Sigma}$ \citep{1999_Allenby_Rossi}, we extend the hierarchical model by 
the conjugate hyperprior distributions
$\phi(\bar{\bm{\upbeta}};\bm{\upmu},\bm{\Omega})$ (normal) and 
$\omega(\bm{\Sigma};\mathcal{v},\bm{\Psi})$ (inverse Wishart) with 
$\mathcal{o}\times1$ mean vector $\bm{\upmu}$, $\mathcal{o}\times\mathcal{o}$
covariance matrix $\bm{\Omega}$, degrees of freedom $\mathcal{v}$ 
$(\geq\mathcal{o})$ and $\mathcal{o}\times\mathcal{o}$ scale matrix $\bm{\Psi}$. 
Using Bayes' theorem and omitting the marginalization over $\bm{\upbeta}_i$, the 
joint posterior of this 3-stage framework can be written as
\begin{gather}
  \begin{gathered}
    \label{posterior}
      P(\bar{\bm{\upbeta}},\bm{\Sigma},\bm{\mathrm{B}}|
        \bm{\mathrm{y}})\propto
          \prod_{i}
            \mathcal{L}(\bm{\mathrm{y}}_i|\bm{\upbeta}_i)
            \phi(\bm{\upbeta}_i|\bar{\bm{\upbeta}},\bm{\Sigma})
            \phi(\bar{\bm{\upbeta}};\bm{\upmu},\bm{\Omega})
            \omega(\bm{\Sigma};\mathcal{v},\bm{\Psi}).
  \end{gathered}
\end{gather}

To obtain draws from $P$, we use a hybrid Gibbs sampler with a random walk 
Metropolis-Hastings step, which allows sampling of $\bm{\upbeta}_i\forall i$ 
alongside the hyperparameters while avoiding simulation of the intractable 
$\mathcal{o}$-dimensional integral from \Cref{mlm} and \Cref{likelihood} 
\citep[see][for details]{2005_Rossi_et_al,2009_Train}. As can be seen in 
\cite{2025_Dressler_et_al}, we refrain from imposing constraints on the 
parameter estimation and always keep the entire Markov chain for evaluation
purposes. However, we control for monotonicity post hoc if necessary (see 
\Cref{sec:modeleval}). In the rest of this paper, we will denote the 
hyperparameters as $\Theta$ and refer to posterior draws of $\bm{\upbeta}_i$ as 
just draws.

\subsection{Practical use}
\label{sec:hbmxlestprac}
\textit{\textbf{Function and input}}
\newline
The corresponding function from \pkg{cash} is \code{E_hbmxlest()}:
\begin{Code}
E_hbmxlest(trainingdesign = "./C_trainingdesign.rds",
  bayesmdata = "./D_bayesmdatatraining.rds", mcmciterations,
  nthiterationtoprint = 5000, moncomponents = 1, nthdrawtokeep = 1,
  signresfirstatt = c("none", "negative", "positive"), compresslist = TRUE,
  seed = 1, temp = TRUE)
\end{Code}

It employs the sampler implemented by the \code{rhierMnlRwMixture()} function 
from the \proglang{R} package \pkg{bayesm} \citep{2025_Rossi}. 

If not specified differently, \code{E_hbmxlest()} automatically imports the 
design and response .rds files from \code{C_} and \code{D_} through its first
two arguments. The length of the Markov chain has to be defined by an integer 
given to \code{mcmciterations}. Pre-set to every 5,000 iterations, the progress
update frequency of the sampler can be controlled via 
\code{nthiterationtoprint}. As a single component approach for the distribution 
of $\bm{\upbeta}_i$ is assumed, the argument \code{moncomponents} may be ignored 
for now. Although we prefer to save the entire chain and avoid constraints in
the estimation, pre-thinning the chain and sign-restricting the draws of the 
$\mathcal{m}-1$ price levels (first feature) can be done with 
\code{nthdrawtokeep} (e.g., \code{nthdrawtokeep = 2} leads to deletion of every 
other draw) and \code{signresfirstatt}, respectively. \code{compresslist} is 
just for reducing the size of the output. 

Following the findings of \cite{2019_Hein_et_al}, we leave the prior 
specifications at default, which is why there are no respective arguments. Here, 
too, the user is invited to make use of the package's or other software's full 
potential by importing her own data into upcoming functions. Compared to
\code{C_} and \code{D_}, no external data preparation is included in \code{E_}. 
As with all the functions in general, the user may look at and mimic the 
structure of the outputs of \code{E_}/inputs of the upcoming functions in such a 
case.
\newpage

\textit{\textbf{Output and example}}
\newline
\code{E_hbmxlest()} only outputs the .rds file
\setlist{nolistsep}
  \begin{itemize}[noitemsep]
    \item \code{E_model.rds},
  \end{itemize}
which is a \code{list} carrying (among other information) the model tensor in 
\code{$betadraw}.

As an example, estimating a model given the (training) design from \code{C_} and 
the responses from \code{D_} can be achieved through
\begin{Schunk}
\begin{Sinput}
R> E_hbmxlest(mcmciterations = 20000, temp = FALSE)
\end{Sinput}
\end{Schunk}
Based on the 50 respondents $\times$ 8 parameters defined in \code{B_} and the
Markov chain length of 20000, the model tensor is of size 50 $\times$ 8 $\times$ 
20000.
\begin{Schunk}
\begin{Sinput}
R> model = readRDS("./E_model.rds")
R> dim(model$betadraw)
\end{Sinput}
\begin{Soutput}
[1]    50     8 20000
\end{Soutput}
\end{Schunk}
Looking at the 19860th and 19960th draw here (the two matrices below) for the 
first two respondents (rows) and first four parameters (columns, price 
coefficients), we can see that in the 19860th draw (top matrix) the first 
respondent's (first row) (unconstrained) utility for the third price level 
(second column) violates the monotonicity introduced to the true preferences in 
\code{B_}. This issue will be addressed in \Cref{sec:modeleval}.
\begin{Schunk}
\begin{Sinput}
R> model$betadraw[1:2, 1:4, c(19860, 19960)]
\end{Sinput}
\begin{Soutput}
, , 1

           [,1]       [,2]      [,3]      [,4]
[1,] -1.2137313 -0.6016924 -3.590299 -4.216515
[2,] -0.6253914 -1.0156327 -1.242151 -3.345381

, , 2

           [,1]      [,2]      [,3]      [,4]
[1,] -2.5411701 -2.639417 -5.133300 -5.384700
[2,] -0.7772189 -1.982588 -2.685474 -3.515665
\end{Soutput}
\end{Schunk}
If \code{$betadraw} is required to hold more than four billion elements, the 
\pkg{RcppArmadillo} error \code{Cube::init(): requested size is too large; 
suggest to enable ARMA_64BIT_WORD} \newline might show up. To resolve this, 
create .R/Makevars (if not already present) in the home directory, add the 
compiler flag \code{CXXFLAGS = -DARMA_64BIT_WORD=1} to the Makevars file and 
reinstall \pkg{bayesm}.
\newpage

\section{Design assessment} 
\label{sec:designeval}
\subsection{Theoretical background}
\label{sec:designevaltheo}
In \Cref{sec:choicedes}, we already mentioned that the design assessment is an 
exception to our naming convention reflecting the order within the ensemble of
functions. The primary reason for this is to be able to presuppose 
\Cref{sec:hbmxlest}'s theoretical underpinnings here as well, thereby avoiding 
redundancy in explanation.

Recall the three well-known principles of efficient choice designs from 
\Cref{sec:choicedes}. We assess them by computing the correlation matrix for 
each design (orthogonality), the marginal frequencies of the feature levels in 
each design (level balance) and the \% of equal feature levels for each pair of 
alternatives within each set within the training and the test designs (minimal 
overlap). As promised, we also calculate the \% of equal alternatives for each 
pair of sets within and between the training and the test designs, allowing to 
check if there exist any set duplicates.

Furthermore, a hierarchical Bayesian mixed logit model (see \Cref{sec:hbmxlest}) 
is estimated based on random responses, which should lead to parameter estimates 
close to zero for an efficient choice design. As a recovery measure, we
determine the root mean squared error (RMSE) between the true (zero) and
re-estimated individual part-worths on parameter level for each draw as well as 
across parameters for each draw. Averaging them across draws then gives us two 
different ARMSE results (parameter level values, total value). The same is done 
with the mean squared error (MSE) and the mean absolute error (MAE).

The statistical criterion that was optimized for in the design generation is not 
re-calculated, as it is usually provided there.

\subsection{Practical use}
\label{sec:designevalprac}
\textit{\textbf{Function and input}}
\newline
The corresponding function from \pkg{cash} is \code{F_designeval()}:
\begin{Code}
F_designeval(trainingdesign = "./C_trainingdesign.rds",
  testdesign = "./C_testdesign.rds", burninpercentage = 0.9, mcmciterations, 
  nthiterationtoprint = 5000, moncomponents = 1, nthdrawtokeep = 1, 
  recoverycheckfortestdesign = TRUE, seed = 1, temp = TRUE)
\end{Code}

The designs are loaded through the first two arguments (if the designs have been
replicated and response columns have been added to prepare external data in 
\code{D_} for model estimation but the design evaluation is of interest 
although the choice experiment must have already taken place, \code{$design} has 
to be changed to single designs without responses). \code{burninpercentage} is 
for discarding burn-in iterations of the Markov chain before calculating the 
recovery measures. Model estimation for the test design based on random 
responses might not be sensible, as it is not used for training, and preventing 
it by setting \code{recoverycheckfortestdesign = FALSE} may even be required to 
avoid Cholesky decomposition errors in the Metropolis-Hastings step if the
matrix has been altered (e.g., by keeping only few sets). The  other arguments 
are identical to \code{E_} (see \Cref{sec:hbmxlest}).
\newline

\textit{\textbf{Output and example}}
\newline
Like \code{E_}, \code{F_designeval()} only outputs a single .rds file, namely
\setlist{nolistsep}
  \begin{itemize}[noitemsep]
    \item \code{F_designeval.rds},
  \end{itemize}
which is a \code{list} containing all the information described above.

Given the defaults and available outputs of previous functions, the call may 
just be
\begin{Schunk}
\begin{Sinput}
R> F_designeval(mcmciterations = 20000, temp = FALSE)
\end{Sinput}
\end{Schunk}
In view of the foregoing definitions, the list elements are mostly 
self-explanatory, but let us peek at the evaluation of our training design. It 
exhibits near orthogonality (lower left/upper right quadrant of correlation
matrix shows small linear dependencies), perfect level balance (equal marginal 
frequencies) and perfect minimal overlap (\code{max overlap within sets} is 
zero). As \code{max overlap between sets} is 60\%, there are no set duplicates.
It is important to mention that, in the field of discrete choice analysis, the 
term (minimal) overlap is usually connected to the comparison of alternatives 
within sets (i.e., the assessment of feature level equality between product 
configurations). Here, we also use overlap to refer to the equality of 
alternatives between sets (necessary to identify set duplicates).
\begin{Schunk}
\begin{Sinput}
R> designeval=readRDS("F_designeval.rds")
R> round(designeval$`orthogonality (correlation matrix) training`, 2)
\end{Sinput}
\begin{Soutput}
      Var12 Var13 Var14 Var15 Var22 Var23 Var24 Var25
Var12  1.00 -0.25 -0.25 -0.25 -0.08  0.08  0.00  0.00
Var13 -0.25  1.00 -0.25 -0.25  0.00 -0.08  0.00  0.00
Var14 -0.25 -0.25  1.00 -0.25  0.00  0.00  0.00  0.08
Var15 -0.25 -0.25 -0.25  1.00  0.00  0.00  0.00  0.00
Var22 -0.08  0.00  0.00  0.00  1.00 -0.25 -0.25 -0.25
Var23  0.08 -0.08  0.00  0.00 -0.25  1.00 -0.25 -0.25
Var24  0.00  0.00  0.00  0.00 -0.25 -0.25  1.00 -0.25
Var25  0.00  0.00  0.08  0.00 -0.25 -0.25 -0.25  1.00
\end{Soutput}
\begin{Sinput}
R> designeval$`level balance (marginal frequencies) training`
\end{Sinput}
\begin{Soutput}
     Var12 Var13 Var14 Var15 Var22 Var23 Var24 Var25
[1,]    15    15    15    15    15    15    15    15
\end{Soutput}
\begin{Sinput}
R> designeval$`overlap within sets training summary`
\end{Sinput}
\begin{Soutput}
     min max mean
[1,]   0   0    0
\end{Soutput}
\begin{Sinput}
R> designeval$`overlap between sets within training summary`
\end{Sinput}
\begin{Soutput}
     min max      mean
[1,] 0.2 0.6 0.3098039
\end{Soutput}
\end{Schunk}
Note that the complete \code{overlap within sets} matrices list every comparison 
of alternatives independent of the resulting \% of equal feature levels (i.e., 
also if two alternatives show 0\% overlap with regard to their feature levels as 
seen below).
\begin{Schunk}
\begin{Sinput}
R> designeval$`overlap within sets training`[1:6, ]
\end{Sinput}
\begin{Soutput}
  set alt1 alt2 overlap
1   1    1    2       0
2   1    1    3       0
3   1    1    4       0
4   1    1    5       0
5   1    2    3       0
6   1    2    4       0
\end{Soutput}
\end{Schunk}
In contrast, the \code{overlap between sets} matrices (check for duplicate sets) 
only show the comparisons leading to at least one match of two alternatives. If 
there are multiple equal alternatives between two sets, all matches except for 
the last one carry \code{NA} in the column \code{overlap}. For example, row 2 of 
the \code{overlap between sets within training} matrix below expresses that 
\code{set 1 alt 1} is equal to \code{set 4 alt 1} (ignore the numbers in the 
column header, as they only serve to indicate that each row is a comparison of 
two alternatives from two sets), but because row 3 provides that 
\code{set 1 alt 3} equal to \code{set 4 alt 4}, row 2 carries \code{NA} and row 
3 0.4. 2 out of 5 alts are equal between those two sets, i.e., 40\%.
\begin{Schunk}
\begin{Sinput}
R> designeval$`overlap between sets within training`[1:3, ]
\end{Sinput}
\begin{Soutput}
  set1 training alt1 training set2 training alt2 training overlap
1             1             5             3             5     0.2
2             1             1             4             1      NA
3             1             3             4             4     0.4
\end{Soutput}
\end{Schunk}

\section{Convergence and model assessment} 
\label{sec:modeleval}
\subsection{Theoretical background}
\label{sec:modelevaltheo}
Guided by the recommendations of \cite{2013_Gelman_et_al} regarding the 
diagnosis of convergence, we simulate an independent second chain with identical
settings except for the seed, discard burn-in iterations, split each chain in 
half, and assess mixing and stationarity by computation of the uni- and 
multivariate potential scale reduction factor ((M)PSRF) \citep{1992_Gelman_Rubin,
1998_Brooks_Gelman} using the within- and between-sequence variances \citep[to 
compute the (M)PSRF, we use the R package \pkg{CODA},][]{2006_Plummer_et_al}. If 
numerous experimental conditions are to be studied, we suggest to 
reverse-engineer a conservative number of iterations to hold constant for 
burn-in and assessment of convergence based on the most demanding treatments
\citep[see][]{2025_Dressler_et_al}.

When monotonicity constraints are imposed on the synthetic utilities, it may be 
reasonable to remove violating draws (incorrect order and signs, see 
\Cref{sec:prefgen}) from the primary chain, and if the lowest remaining number 
across individuals is insufficient, to estimate the necessary primary chain 
length by extrapolation. For each individual, the required number of acceptable 
draws is then collected starting from the end of the (potentially new) primary 
chain. Finally, to reduce serial correlation, the draws should be thinned 
\citep{2009_Train,2013_Gelman_et_al} independently of the removal \citep[for the 
final number of draws to use in simulation, see, e.g.,][]{2000_Orme_Baker}.

In line with the Bayesian estimation, the resulting 
$\mathcal{i}\times\mathcal{o}\times\mathcal{n}$ tensor of draws
$n\in N=\{1,...,\mathcal{n}\}$ is validated by computing credible intervals for 
common measures of parameter recovery and predictive accuracy. If the 
credibility level $1-\alpha$ points at decimal indices for the limits, the 
weighted mean of the two respective values of the measure is taken at both 
limits (i.e., at the lower limit, 
$(\frac{\alpha}{2}\mathcal{n})\,\mathrm{mod}\,1$ and
$1-(\frac{\alpha}{2}\mathcal{n})\,\mathrm{mod}\,1$ as weights for the lower and
upper value, respectively). In the following, we restrict the mathematical 
representations to the draws, but the measures are provided for point estimates 
(posterior means), too.

To assess the parameter recovery, for each draw we calculate the RMSE between 
the true and re-estimated individual part-worths, $\beta_{io}$ and 
$\hat{\beta}_{nio}$, as well as the average Pearson correlation across the 
$\mathcal{i}$ individual part-worth vector pairs, $\bm{\upbeta}_{i}$ and 
$\hat{\bm{\upbeta}}_{ni}$ \citep[see, e.g.,][]{1996_Vriens_et_al,
2002a_Andrews_et_al,2002b_Andrews_et_al,2019_Hein_et_al,2020_Hein_et_al,
2024_Goeken_et_al}. Following \cite{2002a_Andrews_et_al}, we assume that the 
Gumbel's scale $s$, known to be inextricably included in 
$\hat{\bm{\upbeta}}_{ni}$ as divisor of the unscaled part-worths, is estimated 
correctly as the square root of the factor by which the standard variance of 
$\frac{\pi^2}{6}$ is scaled during the error generation in the response 
simulation (see \Cref{MRGEtuning} in \Cref{sec:responsesim}) and can be 
cancelled out by multiplying the estimated part-worths with the true $s$ such 
that\footnote{To be precise, \cite{2002a_Andrews_et_al} introduced the scaling
to the simulated part-worths and re-scaled the RMSE.}
\begin{gather}
  \begin{gathered}
    \label{RMSErec}
      \mathrm{RMSE}_n^{\mathrm{rec}}=
        \sqrt{
          \frac{1}{\mathcal{i}\mathcal{o}}
          \sum_{i,o}\big(s\hat{\beta}_{nio}-\beta_{io}\big)^2
        }.
  \end{gathered}
\end{gather}
In regard to the Pearson correlation, which may be the more appropriate measure 
of parameter recovery given its scaling invariance, we think it is quite 
desirable that the average of coefficients tends to underestimate the true 
correlation. Monte Carlo studies also show that there is a risk of inflating 
positive bias further by employing the Fisher transformation for correction 
\citep[see, e.g.,][]{2015_Bishara_Hittner}. In contrast to previous papers, we 
therefore calculate the mean conservatively as
\begin{gather}
  \begin{gathered}
    \label{corr}
      \mathrm{corr}_n=
        \frac{1}{\mathcal{i}}
        \sum_{i}\mathrm{corr}\big(\hat{\bm{\upbeta}}_{ni},\bm{\upbeta}_{i}\big).
  \end{gathered}
\end{gather}

The out-of-sample predictive accuracy is evaluated through an individual as well
as an aggregate measure for each draw, namely the hit rate across hold-out sets 
averaged over individuals (Equation~\ref{hitrate}) and the shares of (first) 
choice RMSE over alternatives and hold-out sets (Equation~\ref{RMSEsoc}) 
\citep[see, e.g.,][]{1996_Vriens_et_al,2002a_Andrews_et_al,2002b_Andrews_et_al,
2019_Hein_et_al,2020_Hein_et_al,2022_Hein_et_al,2024_Goeken_et_al}.
\begin{gather}
  \begin{gathered}
    \label{hitrate}
      \mathrm{hitrate}_n=
        \frac{1}{\mathcal{i}}
        \sum_{i}\bigg(\frac{1}{\mathcal{k}}\sum_{k,j}\hat{f}_{nijk}f_{ijk}\bigg)
  \end{gathered}
\end{gather}
\begin{gather}
  \begin{gathered}
    \label{RMSEsoc}
      \mathrm{RMSE}_n^{\mathrm{soc}}=
        \sqrt{
          \frac{1}{\mathcal{k}\mathcal{j}}
          \sum_{k,j}
            \bigg(
              \frac{1}{\mathcal{i}}\sum_{i}\hat{f}_{nijk}-
              \frac{1}{\mathcal{i}}\sum_{i}f_{ijk}
            \bigg)^2
        }
  \end{gathered}
\end{gather}
The rationale behind the use of the first choice rule here is rooted in the 
central part of the methodological framework, i.e., \Cref{sec:nashgame}. Given 
the broader scope of required considerations in comparison to the logit choice 
rule, it is just more suitable for the technical explanations, but analogous to 
the parameter sets (draws, means, simulated/true), the 
$\mathrm{RMSE}^{\mathrm{soc}}$ is also reported under the logit rule 
(Equation~\ref{mnl}).

In the above, the focus is not only limited to the draws and first choice rule 
but generally variants of the measures favored by us. To be exhaustive, we 
provide
\setlist{nolistsep}
  \begin{itemize}[noitemsep]
    \item 36 different recovery errors 
          ($3 \times 2 \times 2 \times 2 + 3 \times 2 \times 2 \times 1$):
      \begin{itemize}[noitemsep]
        \item as RMSE, MSE and MAE (3 opts),
        \item on parameter level and in total (2 opts),
        \item with and without cancellation of $s^{-1}$ (2 opts),
        \item as credible interval and draw average (2/1 opt(s), depending on 
              parameter set),
        \item for draws and posterior means (2 opts),
      \end{itemize}
    \item 12 different recovery errors with zero as true parameters for 
          comparison with the measures based on random responses (see 
          \Cref{sec:designeval}) ($3 \times 2 \times 2$):
      \begin{itemize}[noitemsep]
        \item as RMSE, MSE and MAE (3 opts),
        \item on parameter level and in total (2 opts),
        \item for draws and posterior means (2 opts),
      \end{itemize}
    \item 6 different correlations ($2 \times 2 + 2 \times 1$):
      \begin{itemize}[noitemsep]
        \item with and without Fisher transformation (2 opts) (Shapiro Wilk and 
              Kolmogorov Smirnov normality test results for the parameter sets 
              (draws, means, simulated/true) are available in this context),
        \item as credible interval and draw average (2/1 opt(s), depending on 
              parameter set),
        \item for draws and posterior means (2 opts),
      \end{itemize}
    \item 6 different hit rates ($2 \times 2 + 2 \times 1$):
      \begin{itemize}[noitemsep]
        \item for training and test design (2 opts),
        \item as credible interval and draw average (2/1 opt(s), depending on 
              parameter set),
        \item for draws and posterior means (2 opts),
      \end{itemize}
     \item 24 different shares of choice errors 
           ($3 \times 2 \times 2 \times 1 + 3 \times 2 \times 1 \times 2$):
       \begin{itemize}[noitemsep]
        \item as RMSE, MSE and MAE (3 opts),
        \item for training and test design (2 opts),
        \item as credible interval and draw average (2/1 opt(s), depending on 
              parameter set),
        \item for first and logit choice rule (2/1 opt(s), currently logit only 
              for posterior means),
        \item for draws and posterior means (2 opts).
       \end{itemize}
  \end{itemize}

\subsection{Practical use}
\label{sec:modelevalprac}
\textit{\textbf{Function and input}}
\newline
The corresponding function from \pkg{cash} is \code{G_modeleval()}:
\begin{Code}
G_modeleval(experimental = FALSE, 
  simulatedindividualbetas = "./B_simulatedindividualbetas.rds",
  trainingdesign = "./C_trainingdesign.rds", 
  testdesign = "./C_testdesign.rds",
  trainingdesignwithsimuresponses = "./D_trainingdesignwithsimuresponses.rds",
  testdesignwithsimuresponses = "./D_testdesignwithsimuresponses.rds",
  bayesmdata = "./D_bayesmdatatraining.rds", model = "./E_model.rds",
  burninpercentage, chainlengthpsrf, secondmodelforgelman = TRUE,
  seedsecondmodel = 2, removereversals = TRUE, removepositives = TRUE,
  drawspostremoval, thinning, alphanormality = 0.05, crediblelevel = 0.95,
  saveplot = FALSE, show = 5, seed = 1, temp = TRUE)
\end{Code}

The default imports of previous outputs should be clear at this point. If an
internal model based on external data or an external model is to be evaluated, 
set \code{experimental = TRUE}, ignore \code{simulatedindividualbetas} (and 
therefore recovery), let the .rds for \code{trainingdesign} and 
\code{testdesign} be identical to the ones generated by \code{C_} except that 
\code{$design} must be a matrix containing the designs of all respondents with 
responses in the last column (as explained in \Cref{sec:responsesim}). The 
\code{...simuresponses} arguments then have to be .rds comprising just those
matrices (see \Cref{sec:responsesim} as well). The .rds for 
\code{bayesmdata} may be generated with \code{D_} for external data, and 
concerning \code{model} (if external), mimic the output of \code{E_} 
(\code{$nmix} and \code{$loglike} can be disregarded). For model evaluation of 
point estimates only, the matrix has to be duplicated to hand over the tensor 
\code{G_} expects from \code{E_} in \code{$betadraw}.

The next four arguments refer to convergence. If the value of 
\code{burninpercentage} is $<1$, it is interpreted as a percentage, and if it
is $\geq1$, it is interpreted as an absolute number of burn-in iterations to be 
discarded. \code{chainlengthpsrf} serves to define the absolute number of 
iterations after burn-in to be used for computation of the (M)PSRF and
therefore to avoid calculating the latter based on potentially divergent chain 
lengths across multiple experimental conditions. Although not recommended, 
\code{secondmodelforgelman = FALSE} allows to skip simulation of an independent 
second chain, leading to the primary chain being split into two (artificial) 
chains. If an independent second (real) chain shall be simulated, the respective 
seed can be given to \code{seedsecondmodel}. In the latter case, \code{G_} 
always provides the (M)PSRF for split and unsplit chains. If 
\code{saveplot = TRUE}, the last respondent's primary chain trace plots for each 
parameter are saved.

\code{removereversals = TRUE} and \code{removepositives = TRUE} result in the
removal of draws (remaining after burn-in) that violate price monotonicity 
(incorrect order and signs). Note that \code{removepositives} is not considered
if \code{removereversals = FALSE}. The desired number of draws after removal 
before thinning has to be specified via \code{drawspostremoval}. If this 
threshold is not met, \code{G_} stops to execute and tells the lowest remaining 
number across individuals. To resolve this, either \code{drawspostremoval} has 
to be set to this number, a new primary chain has to be simulated with higher 
\code{mcmciterations} (see \Cref{sec:hbmxlest}) or \code{burninpercentage} has 
to be reduced. If the threshold is met, the draws get truncated to
\code{drawspostremoval} and thinned according to \code{thinning} (as with
\code{nthdrawtokeep} in \code{E_}, e.g., \code{thinning = 2} leads to deletion 
of every other draw).

\code{alphanormality} and \code{crediblelevel} take the significance level for 
the frequentist normality tests and the probability coverage of the Bayesian 
credible intervals for the parameter recovery and predictive accuracy measures,
respectively.

Besides the trace plots mentioned above, \code{saveplot} is somewhat similar to 
\code{saveplot} from \code{B_}, but in contrast to the simulated preferences, 
the estimated ones are available as posterior draws and means. Therefore, 
\code{saveplot = TRUE} not only produces a very basic \code{ggplot} of the final 
posterior means with identical settings to the graphic from \code{B_} to 
facilitate their comparison but also visualizes the draws. For each parameter, 
the population's distribution is shown for a selected number of the final 
$\mathcal{n}$ draws in equal distance (controlled by \code{show}, e.g., every 
10th draw if \code{show = 50} and $\mathcal{n} = 500$), as well as the draws' 
distribution for a selected number of respondents.
\newline

\textit{\textbf{Output and example}}
\newline
\code{G_modeleval()} outputs the three .rds files
\setlist{nolistsep}
  \begin{itemize}[noitemsep]
    \item \code{G_modeleval.rds},
    \item \code{G_estimatedindividualbetasdraws.rds} and
    \item \code{G_estimatedindividualbetaspoint.rds}.
  \end{itemize}
The first one is a \code{list} carrying the assessment results and the other two 
are the final preferences arrays to work with.

As an example, let us discard half of the Markov chain (10000 out of 20000 
iterations specified in \code{E_}), use the full other half for the convergence 
checks, set our minimum for the remaining draws after removal of draws violating 
price monotonicity to 500 and the thinning to 5 (every 5th draw is kept).
\begin{Schunk}
\begin{Sinput}
R> G_modeleval(burninpercentage = 10000, chainlengthpsrf = 10000, 
+    drawspostremoval = 500, thinning = 5, temp = FALSE)
\end{Sinput}
\end{Schunk}
This leaves us with a tensor of size 50 $\times$ 8 $\times$ 100 for the draws.
\begin{Schunk}
\begin{Sinput}
R> estimatedindividualbetasdraws = 
+    readRDS("G_estimatedindividualbetasdraws.rds")
R> dim(estimatedindividualbetasdraws)
\end{Sinput}
\begin{Soutput}
[1]  50   8 100
\end{Soutput}
\end{Schunk}
As in \code{F_designeval.rds}, the list elements of \code{G_modeleval.rds} are 
self-explanatory given the information above. Therefore, we will just pick a few 
of them to assess our model, leaving the rest for interested users to explore.

With regard to convergence, the average MPSRF is close enough to perfect 
convergence ($=1$) for our purposes here
\citep[][propose a threshold of 1.1]{2013_Gelman_et_al}.
\begin{Schunk}
\begin{Sinput}
R> modeleval = readRDS("G_modeleval.rds")
R> modeleval$`Gelman mpsrf with real chains split summary`
\end{Sinput}
\begin{Soutput}
       V1       
 Min.   :1.013  
 1st Qu.:1.023  
 Median :1.029  
 Mean   :1.029  
 3rd Qu.:1.034  
 Max.   :1.056  
\end{Soutput}
\end{Schunk}
The highest number of draws per respondent violating price monotonicity is 9077 
out of 10000, which means if we had specified \code{drawspostremoval} > 923, 
\code{G_} would have stopped execution and told us this number via error 
message.
\begin{Schunk}
\begin{Sinput}
R> max(modeleval[[paste0("reversals (and incorrect signs) of price betas ",
+    "sum across draws per resp before removal and thinning")]])
\end{Sinput}
\begin{Soutput}
[1] 9077
\end{Soutput}
\begin{Sinput}
R> #of course accessible via modeleval$`reversals...
R> #[[]] just to be able to break long name correctly
\end{Sinput}
\end{Schunk}
Concerning recovery and predictive accuracy, let us take a look at the 95\%
credible intervals of the last three measures for which the mathematical
representations have been shown in the theoretical background.
\begin{Schunk}
\begin{Sinput}
R> round(modeleval[[paste0("Pearson's r credible interval true=sim vs. ",
+    "draws without Fisher's z trans")]], 3)
\end{Sinput}
\begin{Soutput}
[1] 0.873 0.907
\end{Soutput}
\begin{Sinput}
R> round(modeleval$`hit rate credible interval using draws test`, 3)
\end{Sinput}
\begin{Soutput}
[1] 0.598 0.668
\end{Soutput}
\begin{Sinput}
R> round(modeleval[[paste0("shares of first choice RMSE credible interval ",
+    "using draws test")]], 3)
\end{Sinput}
\begin{Soutput}
[1] 0.056 0.094
\end{Soutput}
\end{Schunk}
To prove that these results indicate good performance of the final model, it is
sensible (aside from theoretical baselines of measures and experience) to 
compare them with results based on a random preference tensor (or any other 
model of your choice). Although not vital here, note that whenever 
\code{removereversals = FALSE} (which has to be the case because if we removed 
reversals, we would have to create a non-violating (non-random) tensor to avoid 
errors), \code{drawspostremoval} is not considered and \code{thinning} has to be 
increased to still be left with the same target number of final draws. In our 
case, \code{drawspostremoval = 500} and \code{thinning = 5} lead to 
$\mathcal{n}=100$. To get the same $\mathcal{n}$ with 
\code{removereversals = FALSE}, specify \code{thinning = 100} to thin 10000 
draws after burn-in down to 100.
\begin{Schunk}
\begin{Sinput}
R> #generate random tensor
R>   set.seed(1)
R>   randomtensor = array(rnorm(50 * 8 * 20000), dim = c(50, 8, 20000))
R> #replace $betadraw in model .rds and save under different name
R>   model = readRDS("E_model.rds")
R>   model$betadraw = randomtensor
R>   saveRDS(model, "E_random.rds")
R> #re-run G_
R>   G_modeleval(model = "E_random.rds", burninpercentage = 10000, 
+      chainlengthpsrf = 10000, removereversals = FALSE, drawspostremoval = 0, 
+      thinning = 100, temp = FALSE)
\end{Sinput}
\end{Schunk}
\begin{Schunk}
\begin{Sinput}
R> estimatedindividualbetasdraws = 
+    readRDS("G_estimatedindividualbetasdraws.rds")
R> dim(estimatedindividualbetasdraws)
\end{Sinput}
\begin{Soutput}
[1]  50   8 100
\end{Soutput}
\begin{Sinput}
R> modeleval = readRDS("G_modeleval.rds")
R> round(modeleval[[paste0("Pearson's r credible interval true=sim vs. ",
+    "draws without Fisher's z trans")]], 3)
\end{Sinput}
\begin{Soutput}
[1] -0.093  0.117
\end{Soutput}
\begin{Sinput}
R> round(modeleval$`hit rate credible interval using draws test`, 3)
\end{Sinput}
\begin{Soutput}
[1] 0.120 0.232
\end{Soutput}
\begin{Sinput}
R> round(modeleval[[paste0("shares of first choice RMSE credible interval ",
+    "using draws test")]], 3)
\end{Sinput}
\begin{Soutput}
[1] 0.200 0.261
\end{Soutput}
\end{Schunk}
The credible intervals of the correlation and hit rate span their baseline
expectations of 0 and 20\% (5 alternatives), respectively, given random 
parameters (i.e., random answers). In contrast to the correlation and hit rate, 
for which a theoretical baseline always exists independent of the actual
preferences, the baseline of the shares of choice error is highly dependent on 
the true distribution of choices (e.g., error will be low if the alternatives 
are chosen equally frequent by the respondents because that is also the result 
of random guessing). As with the first two measures, the shares of choice error
improves drastically when using the original model.

We now have to re-run \code{G_} again with the original model to re-overwrite 
\code{G_}'s outputs which are currently based on the random tensor.
\begin{Schunk}
\begin{Sinput}
R> G_modeleval(burninpercentage = 10000, chainlengthpsrf = 10000, 
+    drawspostremoval = 500, thinning = 5, temp = FALSE)
\end{Sinput}
\end{Schunk}

\section{Pre-computations and Nash competition} 
\label{sec:nashgame}
\subsection{Theoretical background}
\label{sec:nashgametheo}
Before undertaking game-theoretic simulations based on the estimates from 
\Cref{sec:modeleval}, it is necessary to define the competing firms' objective. 
Following the seminal conjoint-based research on non-cooperative competitive 
reactions from the long-run perspective \citep{1993_Choi_DeSarbo,
1997_Green_Krieger}, a firm $w\in W=\{1,...,\mathcal{w}\}$ is assumed to search 
for a product (line) $a\in A=\{1,...,\mathcal{a}\}$ that maximizes the total 
contribution margin $\pi_{wak^-}$ given the partial (excluding $w$) competitive 
scenario $k^-\in K^-=\{1,...,\mathcal{k}^-\}$ by varying price and design (i.e., 
non-price feature levels) simultaneously.

To mathematically formulate and extend this optimization problem, also in view
of the more recent works in this field \citep[see][]{2025_Dressler_et_al}, let 
$J$ now be the set of indices for products in the complete (including $w$) 
competitive scenario $k$ (equivalent to a choice set before), let 
$q\in Q=\{1,...,\mathcal{q}\}$ be the product index in the optimizing firm's 
line such that $Q\subset J$, and take $|Q|(\geq1)$ as exogeneously fixed. 
Furthermore, let $\bm{\mathrm{p}}$ be a $1\times\mathcal{m}$ price vector, and 
let $\bm{\mathrm{c}}$ be a $1\times\mathcal{m}(\mathcal{l}-1)$ vector that 
contains the cost of the non-price feature levels. Because the row vector 
$\bm{\mathrm{x}}_{(\cdot)}$ represents a complete product configuration (see 
\Cref{sec:responsesim}), we are able to describe price and design separately by 
splitting $\bm{\mathrm{x}}_{(\cdot)}$ into the subvectors 
$\bm{\mathrm{x}}_{(\cdot)p}$ and $\bm{\mathrm{x}}_{(\cdot)c}$, respectively. If
we expand $\bm{\mathrm{x}}_{(\cdot)}$ and $\hat{\bm{\upbeta}}_{(\cdot)}$ by the
reference category of each feature beforehand, i.e., change the coding scheme of 
$\bm{\mathrm{x}}_{(\cdot)}$ and add zeros to $\hat{\bm{\upbeta}}_{(\cdot)}$, the 
optimization problem can be written as
\begin{align}
  \label{opt1}
    &\!\max_{{\bm{\mathrm{x}}}_{q\in Q}}&& 
      \pi_{wak^-}=\sum_{q}
        \bigg(\frac{1}{\mathcal{n}}\sum_{n,i}\hat{f}_{niq}\bigg)
        \bigg(
          \bm{\mathrm{p}}\bm{\mathrm{x}}_{qp}^T-
          \bm{\mathrm{c}}\bm{\mathrm{x}}_{qc}^T-
          \delta
        \bigg)\\
    \nonumber
      &&&\xrightarrow{a.s.}\sum_{q}
        \bigg(
          \mathcal{i}
            \iint
              f_{q}(\bm{\upbeta})
              \phi(\bm{\upbeta}|\Theta)
              P(\Theta|\bm{\mathrm{y}})
              d\bm{\upbeta}
              d\Theta
        \bigg)
        \bigg(
          \bm{\mathrm{p}}\bm{\mathrm{x}}_{qp}^T-
          \bm{\mathrm{c}}\bm{\mathrm{x}}_{qc}^T-
          \delta
        \bigg)\\
  \label{opt2}
    &\text{s.t.}&&
      \hat{f}_{niq}=
        \begin{cases}
          1,\,\hat{\bm{\upbeta}}_{ni}\bm{\mathrm{x}}_{q}^T=
              \max_{j'}(\hat{\bm{\upbeta}}_{ni}\bm{\mathrm{x}}_{j'}^T)
              \land |S|=1,\\
          \frac{1}{|S|},\,\hat{\bm{\upbeta}}_{ni}\bm{\mathrm{x}}_{q}^T=
                          \max_{j'}(\hat{\bm{\upbeta}}_{ni}
                            \bm{\mathrm{x}}_{j'}^T)
                          \land |S|>1,\\
          0,\,\text{otherwise},
        \end{cases}\\
  \label{opt3}
    &&&
      S=\{j|\hat{\bm{\upbeta}}_{ni}\bm{\mathrm{x}}_{j}^T=
            \textstyle{\max_{j'}}(\hat{\bm{\upbeta}}_{ni}
              \bm{\mathrm{x}}_{j'}^T)\},\\
  \label{opt4}
    &&&
      \bm{\mathrm{x}}_{q^{*}}\bm{\mathrm{x}}^T_q<\mathcal{l}
      \,\mathrm{if}\,|Q|>1,\\
  \label{opt5}
    &&&
      \bm{\mathrm{x}}_q=(\bm{\mathrm{x}}_{qp},\bm{\mathrm{x}}_{qc}),\\
  \label{opt6}
    &&&
      \sum_{m}x_{qlm}=1,\\
  \label{opt7}
    &&&
      x_{qlm}\in \{0,1\}.
\end{align}
For each product $q$, which has to differ from the other (if existing) products 
$q^{*}\in Q(q^*\neq q)$ in at least one feature (Equation~\ref{opt4}), the 
demand is calculated by taking the sum of first choices $\hat{f}_{niq}$ 
(Equation~\ref{opt2}) over individuals $i$ and averaging it across draws $n$. 
This simulation almost surely converges to the expected demand, which is the 
integral over the distribution of individual preferences $\bm{\upbeta}$ and the 
posterior of hyperparameters $\Theta$ with regard to the first choices (scaled 
by our volume $\mathcal{i}$ here). Compared with logit probabilities, the first 
choices are scaling invariant (i.e., not influenced by the scale factor $s^{-1}$ 
of the part-worths) and immune to the property of independence of irrelevant 
alternatives (i.e., not prone to share inflation for similar products). Their 
possible downside of unrealistic determinism \citep[at least for goods with low 
involvement, see, e.g.,][]{1979_Shocker_Srinivasan} is reduced by implicitly 
obtaining the individual-level share of draws with maximum utility for the 
respective product (note that $\frac{1}{\mathcal{n}}\sum_{n,i}\hat{f}_{niq}$ is 
the same as $\sum_{i}(\frac{1}{\mathcal{n}}\sum_{n}\hat{f}_{niq})$). However, 
$\hat{f}_{niq}$ and $\hat{\bm{\upbeta}}_{ni}$ may be replaced to test the effect 
of various combinations between choice rules (first and logit choice) and 
parameter sets (draws vs. point estimates (posterior means) vs. true 
preferences) on the optimization and equilibria outcomes. We could as well 
incorporate the preference uncertainty by calculating the demand (and optimizing 
the objective function) for every draw of the hyperparameters,
\begin{gather}
  \begin{gathered}
    \label{optintdraw}
      \sum_{i}\hat{f}_{niq}
      \xrightarrow{a.s.}
      \mathcal{i}\int
        f_{nq}(\bm{\upbeta}_n)
        \phi(\bm{\upbeta}_{n}|\Theta_n)
        d\bm{\upbeta}_{n},
  \end{gathered}
\end{gather}
and subsequently simulating the equilibria for every draw like
\cite{2014_Allenby_et_al} to build up posterior distributions of equilibria. 
Though, in our case, the equilibrium quantity is multidimensional \citep[price 
and non-price features here vs. a single metric feature in][]
{2014_Allenby_et_al}, which greatly restricts the manageability of such 
distributions with respect to interpretability and comparability.

If multiple first choices are present within a complete competitive scenario, 
indicated by the cardinality of set $S$, here, the 100\% probability is equally 
divided between them (Equations~\ref{opt2}, \ref{opt3}). Despite being identical 
with high enough frequency, sampling produces slightly different total 
contribution margins for complete competitive scenarios consisting of the same 
set of winning products, which may be unsuitable for the structural analysis of
equilibria (the commonness of recurrences and ties will become apparent in the 
following). It is worth mentioning that we do not (need to) implement such a 
tie-breaking strategy in the response simulation and the model assessment, as a 
tie can only arise there in the highly improbable case of two distinct products
showing the identical and at the same time largest total utility (safety 
mechanism: if this happened in the response simulation, \pkg{bayesm} would throw 
the error \code{nrow(X) ne p*length(yi); exception at unit...}, telling us that 
there are more choices than sets). Furthermore, we do not restrict competition 
by forcing a lower bound on the share of choice 
\citep[cf.][]{2012_Kuzmanovic_Martic}.

The expected demand is then multiplied with the corresponding contribution 
margin of a single unit, which is computed by subtracting $q$'s cost 
$\bm{\mathrm{c}}\bm{\mathrm{x}}_{qc}^T$ as well as a scalar $\delta$ from $q$'s
price $\bm{\mathrm{p}}\bm{\mathrm{x}}_{qp}^T$ \citep[and deliberately allowed to 
be negative in comparison to, e.g.,][]{2019_Kuzmanovic_et_al}. $\delta$ should 
not only comprise a base cost term but also the cost for features that are 
assumed to be unchangeable from the firms' perspective or irrelevant for the 
consumers' choice. Note that the accuracy of the (relative) cost structure most 
definitely is a critical factor affecting the validity of the simulation 
outcomes \citep{1990_Choi_et_al,1994_Choi_DeSarbo}. The marginal cost of 
production is assumed to be constant \citep[see also][]{2014_Allenby_et_al}, as 
is the cost of repositioning.

Due to the discrete domain (i.e., set of binary integers) on which the nonlinear 
objective function (Equation~\ref{opt1}) is defined, the optimization problem is 
of combinatorial nature and a solution cannot be derived analytically. The 
number of theoretically possible product configurations 
$\tau=\mathcal{m}^{\mathcal{l}}$ (continuing the example of symmetric designs) 
and the number of theoretically possible product line configurations 
$\mathcal{a}=\binom{\tau}{\mathcal{q}}$ grow exponentially with $\mathcal{l}$ 
and $\mathcal{q}$, respectively. As is the case with numerous combinatorial 
optimization problems, there also exists no exact numeric algorithm capable of 
solving it in polynomial time, making it NP-hard \citep{1990_Kohli_Sukumar}. In
line with the effort of eliminating uncontrolled systematic influences, we 
nevertheless optimize using complete enumeration such that the structural 
properties of the equilibria can be fully captured and are not biased by 
artefacts from heuristics. 

Apart from that, complete enumeration's time complexity does not always have to 
be disadvantageous, which is why we also refrain from implementing other exact 
methods for now. Compared with procedures guaranteeing a solution's global 
optimality, even the full potential of heuristics \citep[see][for overviews]
{2008_Belloni_et_al,2024_Baier_Voekler} in terms of runtime superiority does not 
come into play when the solution space $A$ of the optimization problem is rather 
small. In the simulation of competitive reactions the latter can be 
computationally very limited if there is a large number of optimization problems 
to be solved consecutively. This will be elaborated upon in the subsequent.

Given the discrete domain of the objective function, our game-theoretic solution 
concept of interest cannot be derived analytically either. Hence, we simulate
dynamic closed-loop games (i.e., multi-stage games with mutually observable past
actions) of myopic best responses in a sequential manner to obtain the fixed
points known as pure strategy Nash equilibria \citep{1838_Cournot,1951_Nash,
1991_Fudenberg_Tirole}.

More precisely, and again closely following \cite{1993_Choi_DeSarbo} and
\cite{1997_Green_Krieger} as well as \cite{1995_Gutsche} and 
\cite{2000_Steiner_Hruschka}, firms take turns in maximizing their total 
contribution margin $\pi_{wak^-}$ depending on the others' product lines until 
no firm can benefit from unilaterally changing its product line. In accordance 
with the aforementioned articles, the competitors are all assumed to be active 
and to be symmetric in regard to prices, cost structure, estimated consumer 
preferences and number of products. Moreover, the number of competitors is 
expected to remain constant throughout a game \citep[see also][]
{2014_Allenby_et_al}.

If the sequence of each firm optimizing once, 
\begin{gather}
  \begin{gathered}
    \label{optseq}
      \max_a\pi_{wak^-}\forall w,
  \end{gathered}
\end{gather}
is denoted a round $b \in B=\{1,...,\mathcal{b}\}$ and $k^-_0\in K^-$ is the
index of the initial state (i.e., the partial competitive scenario at the 
beginning of the $\mathrm{t\hat{a}tonnement}$), here, a Nash equilibrium can be 
formally expressed as the singleton
\begin{gather}
  \begin{gathered}
    \label{nasheq}
      \mathcal{T}^{k^-_0}=
        \{k^{k^-_0}_{b}|k^{k^-_0}_{b}=k^{k^-_0}_{b-1},b\geq 2\}.
  \end{gathered}
\end{gather}
$\mathcal{T}^{k^-_0}$ contains the index $k^{k^-_0}_{b}$ referencing the 
complete competitive scenario $k$ which is present at the end of round $b$ and 
(first) remained unchanged for two consecutive rounds after starting from 
$k^-_0$. It is crucial to set an upper limit for the rounds in order to prevent
a game from running infinitely in the absence of an equilibrium, especially if
no comprehensive detection mechanism is implemented that checks for all
different types of cycles. The latter requires the (partial) comparison between
the current $k_{b}$ and each of the preceding $k_{b'}$ with 
$b' \in B'=\{1,...,b-2\}$. We decided to just look for the shortest possible
cycle (2-round cycle) and this only once when the upper limit of the rounds is 
reached (this is more efficient if the expected number of cyclic games is rather 
low because then this examination is not even triggered once in the majority of 
games),
\begin{gather}
  \begin{gathered}
    \label{tworoundcycle}
      \mathcal{c}^{k^-_0}_2=
        \begin{cases}
          1,\,k^{k^-_0}_{\mathcal{b}}=k^{k^-_0}_{\mathcal{b}-2},\\
          0,\,\text{otherwise}.
        \end{cases}
  \end{gathered}
\end{gather}
On the other hand, at least two rounds have to be played to see if 
$\mathcal{T}^{k^-_0}\neq\emptyset$ because regardless of the order of movement 
of the firms, there is no initial state $k_0$ to meaningfully compare $k_1$
with. In other words, a computational dependence on an initial product line
configuration of the firm that comes first in the reaction sequence does not 
exist (and we do not choose one at random), as each game starts with this firm 
optimizing over its $\mathcal{a}$ possibilities.

Thus, there are at most $\mathcal{k}^-$ initial states 
($\mathcal{a}^\mathcal{w-1}$ theoretically possible competitive scenarios of
$\mathcal{w}-1$ firms) to start a game from, which we exhaustively go through
for two reasons. Firstly, the effect of each unique initial competitive scenario
as well as its order of movement variants (thanks to symmetric competitors) on
the equilibria can be observed, and secondly, every existing equilibrium is
guaranteed to be found since they are inevitably represented in the initial
states.

It is noteworthy that with complete enumeration the firm reacting first already 
needs to calculate $\pi_{wak^-}$ once for all $\mathcal{k}$ (i.e., 
$\mathcal{a}^\mathcal{w}$) theoretically possible complete competitive scenarios 
when optimizing over its $\mathcal{a}$ product line configurations at the start
of each of the $\mathcal{k}^-$ games. If enough memory is available, there are 
major advantages to pre-computing a matrix 
\vspace*{0.2cm}

\AtBeginEnvironment{bNiceMatrix}{\everymath{\displaystyle}}
\begin{gather}
  \begin{gathered}
    \label{m}
      \bm{\mathcal{M}}=
        \begin{bNiceMatrix}
          a_{11}
            &a_{21}
            &\dots
            &a_{\mathcal{w}1}
            &\pi_{11}\\
          \vdots&
            \vdots&
            \ddots&
            \vdots&
            \vdots\\
          a_{1\mathcal{k}}
            &a_{2\mathcal{k}}
            &\dots
            &a_{\mathcal{w}\mathcal{k}}
            &\pi_{1\mathcal{k}}\\
          \CodeAfter
            \OverBrace[yshift=1.5mm,shorten]{1-1}{3-4}
              {\scriptsize{\text{complete competitive scenario}}}
        \end{bNiceMatrix}
      \in\mathbb{R}^{\mathcal{k}\times(\mathcal{w}+1)}
  \end{gathered}
\end{gather}
comprising the $\mathcal{k}$ complete competitive scenarios as well as the 
corresponding total contribution margins $\pi_{wk}\forall k$ (i.e., 
$\pi_{wak^-}\forall a,k^-$) from an arbitrary but constant viewpoint $w$ (e.g., 
$w=1$). After pre-optimizing $\mathcal{k}^-$-times over $\mathcal{a}$ product 
line configurations in $\bm{\mathcal{M}}$, the resulting matrix

\begin{gather}
  \begin{gathered}
    \label{mopt}
      \bm{\mathcal{M}}^{\mathrm{opt}}=
        \begin{bNiceMatrix}
          \argmax_a\pi_{1a1}
            &a_{21}
            &\dots
            &a_{\mathcal{w}1}
            &\max_a\pi_{1a1}\\
          \vdots&
            \vdots&
            \ddots&
            \vdots&
            \vdots\\
          \argmax_a\pi_{1a\mathcal{k}^-}
            &a_{2\mathcal{k}^-}
            &\dots
            &a_{\mathcal{w}\mathcal{k}^-}
            &\max_a\pi_{1a\mathcal{k}^-}\\
          \CodeAfter
            \OverBrace[yshift=1.5mm,shorten]{1-2}{3-4}
              {\scriptsize{\text{partial competitive scenario}}}
        \end{bNiceMatrix}
      \in\mathbb{R}^{\mathcal{k}^-\times(\mathcal{w}+1)}
  \end{gathered}
\end{gather}
can easily be utilized as a look-up table for the best response 
$\argmax_a\pi_{wak^-}$ to a given partial competitive scenario $k^-$. Recalling 
the property of symmetry, it is evident that $\bm{\mathcal{M}}^{\mathrm{opt}}$ 
allows to circumvent the repeated calculation of identical total contribution 
margins and optima across firms and rounds in the games. Consequently, the 
runtime of the $\mathcal{k}^-$ games to be played in an experimental condition 
becomes neglectable. Since increasing compute is usually less feasible than 
increasing memory, we chose to take this memory-heavy path.\footnote{For the 
large-scale Monte Carlo study in \cite{2025_Dressler_et_al}, we used three rack 
servers with 104 physical cores and 2,560 GiB RAM in total:
1x Dell\textsuperscript{\texttrademark} 
   PowerEdge\textsuperscript{\texttrademark} R450 with 
2x Intel\textsuperscript{\textregistered}
   Xeon\textsuperscript{\textregistered} Silver 4316 CPUs and
16x 64 GiB DDR4 3200 MT/s RDIMMs,
2x Dell\textsuperscript{\texttrademark} 
   PowerEdge\textsuperscript{\texttrademark} R440s with 
2x Intel\textsuperscript{\textregistered} 
   Xeon\textsuperscript{\textregistered} Silver 4216 CPUs and
12x 64 GiB DDR4 3200 MT/s RDIMMs each.}

As implicitly stated above, the $\mathcal{k}^-$ optimization problems to be 
solved ($\mathcal{k}^-\mathcal{w}\mathcal{b}$ without pre-computation) pose the 
main computational bottleneck. Due to the exponential growth with $\mathcal{w}$, 
the solution space $A$ of a single optimization problem can be so small that 
both heuristics and exact methods are comparably fast but the vast initial state 
space $K^-$ precludes computational feasibility. Thus, if $A$ calls for 
heuristics, $K^-$ certainly will too. The primary way of restoring feasibility 
therefore is to prune $K^-$. To the best of our knowledge, such an approach has 
not yet been developed, but even if it were to exist, it would only be employed 
here with the presence of a mathematical proof demonstrating the equivalence of 
results.

Apart from the latter, the runtime of the pre-computation of $\bm{\mathcal{M}}$
is substantially decreased by implementing the workhorse functions in C++
\citep[\pkg{Rcpp},][]{2026_Eddelbuettel_et_al}, and fortunately, the task is 
embarrassingly parallel (for reproducible parallel computing, we use the 
\proglang{R} packages \pkg{doParallel} \citep{2022a_Microsoft_Weston}, 
\pkg{foreach} \citep{2022b_Microsoft_Weston} and \pkg{doRNG} 
\citep{2026_Gaujoux}). Additionally, we pre-compute the possible line and 
product configurations of a firm, the $\tau\times\mathcal{i}\times\mathcal{n}$ 
tensor of (exponentiated) product utilities as well as the $\tau\times 1$ vector 
of product contribution margins to serve as look-ups \citep{2008_Belloni_et_al}. 
This avoids unnecessary re-computations also within the main pre-computation 
process of $\bm{\mathcal{M}}$ itself. With regard to memory, the number of 
elements in $\bm{\mathcal{M}}^{({\mathrm{opt}})}$ is minimized by mapping the 
pre-computed extended line and product configurations of each firm to a single 
integer analogous to \Cref{m}.

We were given the opportunity to test our implementation (modified to GPU 
computing) on a blade server of an exascale supercomputer currently under 
development. Note that an extrapolation of the results indicated that even on 
entire machines leading the TOP500 list, simulation of many interesting, yet
moderate scenarios would still be far out of reach, which casts a different 
light on computational limitations the user might encounter on usual machines.

\subsection{Practical use}
\label{sec:nashgameprac}
\textit{\textbf{Function and input}}
\newline
The corresponding functions from \pkg{cash} are \code{H_precomputeM()}:
\begin{Code}
H_precomputeM(trainingdesign = "./C_trainingdesign.rds",
  estimatedindividualbetas = "./G_estimatedindividualbetasdraws.rds",
  pricecostmatrix, pricecostmatrixfull, basecostfix = 0, basecostvar = TRUE,
  choicerule = c("first", "logit"), 
  ruletosolvefcutilitytie = c("splitting", "sampling"), prodsperline,
  competitors, addnullparameters = TRUE, progressupdatedelay = 120, 
  compressmatrix = FALSE, seed = 1, temp = TRUE)
\end{Code}
and \code{I_preoptandnashgame()}:
\begin{Code}
I_preoptandnashgame(scenariosmatrix = "./H_scenariosmatrixfirstdraws.rds",
  scenariosinfo = "./H_scenariosinfofirstdraws.rds",
  products = "./H_products.rds", lines = "./H_lines.rds", maxrounds = 20,
  progressupdatedelay = 120, temp = TRUE)
\end{Code}

As its name reveals, \code{H_precomputeM()} is responsible for the 
pre-computation of $\bm{\mathcal{M}}$, which makes it the computationally most 
demanding function in \pkg{cash} (depending on the settings, of course). If it 
is to be used independently (i.e., with external data), it may be helpful to 
know that from \code{C_trainingdesign.rds} it only extracts \code{$resps}, 
\code{$betas} and \code{$lvls}. 

Its argument \code{estimatedindividualbetas} takes the final preference array of 
draws, point estimates/posterior means or simulated/true utilities.

Price and cost levels have to be imported from two .csv files through 
\code{pricecostmatrix} (a .csv only comprising the features that have been 
defined in \code{B_} as being part of the conjoint choice experiment) and
\code{pricecostmatrixfull} (a .csv containing all features relevant to the 
contribution margin). These .csv have to have row names as first column, column 
names as first row, as many rows as there are features + 1, as many columns as 
there are (max) levels + 1, and price as first feature. Undefined cells (if the
number of levels differs across features) must be \code{NA}, and a zero 
indicates that the level has no cost. Obviously, the order of the features and
levels has to match the order of the estimates in the preference array, e.g., 
the price levels have to be sorted in ascending order if a monotonicity 
constraint was introduced to the price feature in \code{B_} and enforced in 
\code{G_}. Via \code{basecostfix}, the fixed part of the base cost term $\delta$ 
can be set, while \code{basecostvar = TRUE} leads to the levels of the features 
included in the full but not partial price cost matrix being randomly sampled to 
then add their corresponding costs to $\delta$. If $\delta$ shall only be a 
fixed term ($\geq0$), the same .csv can be given to both price cost matrix 
arguments, but \code{basecostvar} must be \code{FALSE}.

If \code{"first"} as the default of \code{choicerule} is to be applied, 
\code{ruletosolvefcutilitytie} allows to choose between \code{"splitting"} and 
\code{"sampling"} as a tie-breaking strategy for the case of multiple first 
choices being present within a complete competitive scenario. 
\code{prodsperline} and \code{competitors} both have to be given an integer 
specifying the number of products per line and number of competitors, 
respectively.

\code{addnullparameters = FALSE} suppresses the expansion of the preference 
array by zeros for the reference category, which may be necessary in case of an
external preference array. The progress of the pre-computation of 
$\bm{\mathcal{M}}$ is saved to a .txt file. Its update frequency can be 
controlled via a numeric value (minutes) given to \code{progressupdatedelay}.

\begin{tcolorbox}[width=0.95\textwidth,halign=justify,center,leftrule=2mm,
                  rounded corners]
  \textbf{Important note}: The first feature in the price cost matrices must
  be price. Entries have to be \code{NA} if a level does not exist and zero if 
  it has no cost. The order of the estimates must match the order of the 
  features and levels in the price cost matrices (e.g., ascending order for the 
  prices if monotonicity was introduced in \code{B_} and enforced in \code{G_}). 
  The maximum number of competitors is five.
\end{tcolorbox}

\code{I_preoptandnashgame()}, on the other hand, pre-computes
$\bm{\mathcal{M}}^{{\mathrm{opt}}}$ and uses it for fast simulation of the Nash 
games. 

To do so, it needs $\bm{\mathcal{M}}$ from \code{H_}, as well as additional 
information to, e.g., convert the single-integer line-identifiers back to 
readable product feature levels. The corresponding .rds files for these first 
two arguments (\code{scenariosmatrix} and -\code{info}) differ in results and 
name depending on the choice rule and parameter settings (so, the defaults may 
need to be changed). The other two, which contain the pre-computed product and 
line configuration matrices, do not. \code{maxrounds} defines the the upper 
limit $\mathcal{b}$ of the game rounds and has to be an even integer (for 
internal reasons).
\newline

\textit{\textbf{Output and example}}
\newline
\code{H_precomputeM()} outputs the five .rds files
\setlist{nolistsep}
  \begin{itemize}[noitemsep]
    \item \code{H_lines.rds},
    \item \code{H_products.rds},
    \item \code{H_scenariosinfo[choice rule][parameter set].rds},
    \item \code{H_scenariosmatrix[choice rule][parameter set].rds} and
    \item \code{H_scenariosmatrixhead[choice rule][parameter set].rds}.
  \end{itemize}
We have already explained the first four above, as they serve as inputs for 
\code{I_}. The last one holds the first 10 rows of $\bm{\mathcal{M}}$ for quick 
checks.
\newpage

\code{I_preoptandnashgame()} only outputs the .rds file
\setlist{nolistsep}
  \begin{itemize}[noitemsep]
    \item \code{I_equilibriainfo[choice rule][parameter set].rds},
  \end{itemize}
which includes all relevant information about the equilibrium simulations. Note 
that the terminal \code{A-I_finaloutput.rds} also gets this choice rule and 
parameter set name extension.

Both functions additionally generate a similar \code{[]_}progress.txt file.

As an example, we let \code {H_} import the final preference array of draws from 
\code{G_}, which is the default, and the following two .csv, pcm2atts.csv and 
pcm6atts.csv, with the price and cost levels for \code{pricecostmatrix} (recall 
that we defined 2 features with 5 levels each in \code{B_}) and 
\code{pricecostmatrixfull} (6 features with 5 levels each), respectively:
\begin{Code}
lvl1,lvl2,lvl3,lvl4,lvl5    | lvl1,lvl2,lvl3,lvl4,lvl5
price,299,599,899,1199,1499 | price,299,599,899,1199,1499
displaysize,25,30,33,44,54  | displaysize,25,30,33,44,54
                            | cpu,10,11,12,65,79
                            | ssdstorage,11,11,11,23,31
                            | batterylife,8,8,10,10,12
                            | ram,6,6,9,19,38
\end{Code}
These price and cost levels (in €) are taken from \cite{2025_Dressler_et_al}, 
where we were able to gain insight into the (relative) cost structure of a
well-known computer manufacturer and modeled competition between firms in this 
sector. In addition to price, we determined five modifiable, discriminating 
features driving the consumers' choice and the firms' contribution margin for
notebooks, namely display size, central processing unit (CPU), solid state drive 
(SSD) capacity, battery life and random access memory (RAM). For the purposes of
our example, further elaboration on the study's details is not necessary. 
Specific feature levels will be provided as needed.

Let us leave the two base cost arguments at preset, which means ignoring the 
fixed part of $\delta$ (no cost for, e.g., housing and mainboard, and no 
explicitly chosen cost levels for CPU, SSD capacity, battery life and RAM) but 
sampling cost levels for the CPU, SSD capacity, battery life and RAM from the 
full price cost matrix. Let us also use the defaults for the choice rule, i.e.,
first choice, as well as the tie-breaking strategy, i.e., splitting, and 
simulate competition between 2 firms, each offering a product line consisting of
2 products. \code{progressupdatedelay} is set to a few seconds to force 
H\code{_}progress.txt to be generated.

As the .rds defaults in \code{I_} correspond to the first choice and draws 
settings employed here in \code{H_}, there is no need to adjust any file names
when calling \code{I_} in this case.
\begin{Schunk}
\begin{Sinput}
R> H_precomputeM(pricecostmatrix = "./pcm2atts.csv", 
+    pricecostmatrixfull = "./pcm6atts.csv", prodsperline = 2, 
+    competitors = 2, progressupdatedelay = 0.15, temp = FALSE)
R> I_preoptandnashgame(temp = FALSE)
\end{Sinput}
\end{Schunk}

Because \code{I_equilibriainfo[][].rds} comprises all contents of 
\code{H_scenariosinfo[][].rds} and more, we will skip the latter and jump 
directly to the former and its most important components. But before we do so, 
it seems sensible to practically explain the structure of $\bm{\mathcal{M}}$ and
the progress file by taking a look at \code{H_scenariosmatrixhead[][].rds} and 
H\code{_}progress.txt.
\begin{Schunk}
\begin{Sinput}
R> scenariosmatrixheadfirstdraws=
+    readRDS("H_scenariosmatrixheadfirstdraws.rds")
R> scenariosmatrixheadfirstdraws
\end{Sinput}
\begin{Soutput}
      [,1] [,2]     [,3]
 [1,]    1    1 5629.350
 [2,]    1    2 2422.785
 [3,]    1    3 7645.995
 [4,]    1    4 5216.270
 [5,]    1    5 7656.210
 [6,]    1    6 7656.210
 [7,]    1    7 3688.400
 [8,]    1    8 7653.940
 [9,]    1    9 6665.180
[10,]    1   10 7656.210
\end{Soutput}
\end{Schunk}
Like \Cref{m} showed theoretically, $\bm{\mathcal{M}}$ consists of as many 
columns as there are competitors plus one for the contribution margin. The 
integers in the competitors' columns refer to rows in the pre-computed line 
configuration matrix which contain integers referring to rows in the 
pre-computed product configuration matrix which contain interpretable feature 
levels. The unit of the contribution margin is defined by the unit of the price
cost matrix entries.

The progress file reports the following information:
\begin{Code}
line competitor 1: [finished #lines/total #lines]
line competitor 2: [finished #lines/total #lines]
start time: [start time and date of computing the matrix]
current time: [current time and date]
elapsed time: [time difference in minutes]
current iterations: [finished #scenarios, i.e., finished #rows of the matrix]
total iterations: [total #scenarios, i.e., total #rows of the matrix]
estimated total time: [in minutes]
estimated remaining time: [in minutes]
\end{Code}
If the specified delay is very short, some oscillation of the time estimates is 
normal due to performance fluctuation across cores in parallelization.

Now, we finally arrive at what we have been working towards throughout this 
paper and package, namely the output of \code{I_}. Its \code{list} begins with 
the provision of a few metrics describing $\bm{\mathcal{M}}$ and 
$\bm{\mathcal{M}}^{{\mathrm{opt}}}$. Having defined 2 features with 5 levels 
each, the number of theoretically possible product configurations is 25. 
\begin{Schunk}
\begin{Sinput}
R> equilibriainfofirstdraws=readRDS("I_equilibriainfofirstdraws.rds")
R> equilibriainfofirstdraws$`number of products`
\end{Sinput}
\begin{Soutput}
[1] 25
\end{Soutput}
\end{Schunk}
By choosing competition on a product line with 2 products, we get 300 
theoretically possible line configurations ($\binom{25}{2}$, i.e., 25 $\times$ 
25 minus 25 lines of identical products, divided by two to get rid of the flips).
\begin{Schunk}
\begin{Sinput}
R> equilibriainfofirstdraws$`number of lines`
\end{Sinput}
\begin{Soutput}
[1] 300
\end{Soutput}
\end{Schunk}
Specifying 2 competitors then leads to 90,000 theoretically possible complete 
competitive scenarios, i.e., 90,000 contribution margins to be pre-computed 
(rows of $\bm{\mathcal{M}}$) and 300 optimization problems to be pre-solved 
(rows of $\bm{\mathcal{M}}^{{\mathrm{opt}}}$).
\begin{Schunk}
\begin{Sinput}
R> equilibriainfofirstdraws$`number of scenarios`
\end{Sinput}
\begin{Soutput}
[1] 90000
\end{Soutput}
\end{Schunk}

Furthermore, we can access the sampled variable base cost levels (matching the
order of the (excess) features in the price cost matrix).
\begin{Schunk}
\begin{Sinput}
R> equilibriainfofirstdraws$`variable base cost`
\end{Sinput}
\begin{Soutput}
[1] 10 23  8  6
\end{Soutput}
\end{Schunk}

Extending the \code{[]_}progress.txt, both functions output a run time and 
memory usage matrix. While for \code{I_}, it reports just the run time (mins) 
and object size (Mb) for different parts of the function, for \code{H_}, it also
calculates the percentage of free system memory taken up (currently only on 
Windows and Linux).
\begin{Schunk}
\begin{Sinput}
R> equilibriainfofirstdraws$`run time and memory usage scenarios`[1:2]
\end{Sinput}
\begin{Soutput}
                                        run time (mins) object size (Mb)
data, parallelism, checks, etc.                   0.010            0.000
add null parameters                               0.000            0.382
compute products                                  0.000            0.000
compute contribution of products                  0.000            0.000
parallel compute utility of products              0.049            0.954
compute lines                                     0.000            0.002
parallel compute scenarios/contribution           0.183            2.086
execute function                                  0.241          181.251
\end{Soutput}
\end{Schunk}

With regard to the equilibria, \code{I_} presents information about them in 
three primary ways. As there are $\mathcal{k}^-$ initial states 
($\mathcal{a}^\mathcal{w-1}$ theoretically possible (partial) competitive 
scenarios of $\mathcal{w}-1$ firms) to start a game from (here, 300), identical 
equilibria may arise multiple times. Such duplicates are generally only counted 
once per run (i.e., per function call). Flips, which are equilibria consisting 
of the same products or lines but swapped between firms (recall the competitors’ 
symmetry), are not treated as duplicates. Note that flips are only possible for 
differentiated equilibria, i.e., equilibria in which not all competitors share 
the same product (line).

In the following first matrix, the equilibria are listed in form of their 
single-integer line-identifiers, i.e., as rows of $\bm{\mathcal{M}}$. Identical 
equilibria are included once except if the number of rounds to reach them 
differs between initial states (there exists a reduced form in 
\code{$`unique equilibria in lines`} without rounds and therefore without 
duplicates due to round differences).

It also saves the initial states (instead of the last product lines present) 
from which no equilibrium has been reached for further investigation. If a game 
in a run does not end in an equilibrium, either more rounds have to be played or 
a cycle is present. As written in the theoretical background, we are only 
certain about the 2-round cycles. To distinguish between these two cases, the 
first competitor (first column) is \code{NA} if a 2-round cycle has been 
detected and zero otherwise (larger cycle present or round limit too low).

In our example, three equilibria have been found. \code{54-51} is a flip of 
\code{51-54}, and \code{130-130} shows up three times because it took a 
different number of rounds to reach it from some of the underlying initial 
states. There exist no initial states from which no equilibrium was reached. 
Otherwise, there would have been rows with \code{NA} or \code{0} in column 
\code{linecomp1} and corresponding initial states (instead of the last product 
lines present) in column \code{linecomp2}. \code{130-130} is the pareto-superior 
equilibrium, as it results in much higher contribution margins for both 
competitors.
\begin{Schunk}
\begin{Sinput}
R> equilibriainfofirstdraws$`unique equilibria in lines with rounds`
\end{Sinput}
\begin{Soutput}
  linecomp1 linecomp2 contribution rounds
1       130       130     12989.16      2
2        51        54      6792.06      1
3       130       130     12989.16      1
4        54        51      6570.30      1
5       130       130     12989.16      3
\end{Soutput}
\end{Schunk}

\begin{tcolorbox}[width=0.95\textwidth,halign=justify,center,leftrule=2mm,
                  rounded corners]
  \textbf{Important note}: If equilibria are found, the underlying initial 
  states are (currently) not listed because from a managerial standpoint, the 
  equilibria themselves are to inform the product strategy (e.g., select the 
  most profitable one), rather than the corresponding initial states leading to 
  them. This perspective also reinforces our approach of considering the entire 
  initial state space (even though, in practice, a firm faces just one 
  competitive scenario at any given time), as well as exact optimization 
  procedures to avoid any loss of equilibrium information.
\end{tcolorbox}

The second matrix contains as many rows as there are (non-duplicate) equilibria
plus as many rows as necessary to view each of those equilibria from the 
perspective of each competitor to calculate the corresponding contribution 
margins. In \code{130-130}, there is no differentiation between the competitors, 
which leaves us with the same contribution margin for both of them (rows 1, 2). 
In \code{51-54}, on the other hand, the two competitors differ in their product 
lines, leading to two different contribution margins (rows 3, 4). Note that the 
column names have to be ignored, as the lines of each equilibrium get exchanged 
to be able to always treat the first column as the competitor from whose 
perspective the contribution margin is calculated. E.g., in the original second 
equilibrium \code{51-54} (row 3), competitor 2 chooses line 54. In row 4, we act 
as if competitor 1 chooses line 54 to be able to calculate the contribution 
margin the same way. In row 4, competitor 1 is therefore to be interpreted as 
competitor 2. The two resulting rows for the flip of \code{51-54} may be 
ignored, as they do not provide any additional insight.
\begin{Schunk}
\begin{Sinput}
R> equilibriainfofirstdraws$`contribution for each competitor`
\end{Sinput}
\begin{Soutput}
  linecomp1 linecomp2 contribution
1       130       130     12989.16
2       130       130     12989.16
3        51        54      6792.06
4        54        51      6570.30
5        54        51      6570.30
6        51        54      6792.06
\end{Soutput}
\end{Schunk}

Ultimately, the third way of presenting the equilibria is a list of matrices in 
which the equilibria are converted back to a readable feature level format. Each 
matrix has as many rows as there are products per line times competitors and as 
many columns as there are product features. In our example, each equilibrium 
consists of 2 products with 2 features for each of the 2 competitors.
\begin{Schunk}
\begin{Sinput}
R> equilibriainfofirstdraws$`unique equilibria in levels`
\end{Sinput}
\begin{Soutput}
[[1]]
           att1 att2
comp1prod1    2    2
comp1prod2    2    3
comp2prod1    2    2
comp2prod2    2    3

[[2]]
           att1 att2
comp1prod1    1    3
comp1prod2    2    2
comp2prod1    1    3
comp2prod2    2    5

[[3]]
           att1 att2
comp1prod1    1    3
comp1prod2    2    5
comp2prod1    1    3
comp2prod2    2    2
\end{Soutput}
\end{Schunk}
Through comparison of the numbers with the actual feature levels \citep[see][for 
the detailed feature level table complementing the price cost matrix]
{2025_Dressler_et_al}, we can conclude that in \code{130-130} (first matrix) 
both firms offer two notebooks at 599 € (second price level), one with a 14" and
one with a 15" display (second and third display level). 
\begin{Code}
Matrix 1 (130-130):
comp1: 599€ 14" + 599€ 15"
comp2: 599€ 14" + 599€ 15"
\end{Code}
In \code{51-54} and its flip (second and third matrix) both competitors offer 
one notebook at 299 € (first price level) with a 15" display (third display 
level). For the second product in their lines, both offer a notebook at 599 € 
(second price level), but one offers it with a 14" display (second display 
level) and the other with a 17" display (fifth display level).
\begin{Code}
Matrix 2 (51-54):
comp1: 299€ 15" + 599€ 14"
comp2: 299€ 15" + 599€ 17"
\end{Code}

\section{Market share and revenue as special cases of profit optimization} 
\label{sec:sharerevopt}
Due to the fact that market share and revenue are both implicitly embedded in 
the objective function of the profit optimization problem presented herein 
(Equation~\ref{opt1} in Section \ref{sec:nashgame}),
\begin{gather}
  \begin{gathered}
    \label{optprof}
      \sum_{q}
        \bigg(\frac{1}{\mathcal{n}}\sum_{n,i}\hat{f}_{niq}\bigg)
        \bigg(
          \bm{\mathrm{p}}\bm{\mathrm{x}}_{qp}^T-
          \bm{\mathrm{c}}\bm{\mathrm{x}}_{qc}^T-
          \delta
        \bigg),
  \end{gathered}
\end{gather}
there is no need to alter the framework to be able to simulate corresponding 
equilibria. For competition on market share, prices and costs must not enter the 
objective function. If the price vector $\bm{\mathrm{p}}$ and the cost vector 
$\bm{\mathrm{c}}$ are replaced with the all-ones vector $\mathbb{1}$ and the 
null vector $\mathbb{0}$, respectively, and the base cost $\delta$ is set to 
zero as well,
\begin{gather}
  \begin{gathered}
    \label{optshare}
      \sum_{q}
        \bigg(\frac{1}{\mathcal{n}}\sum_{n,i}\hat{f}_{niq}\bigg)
        \bigg(
          \mathbb{1}\bm{\mathrm{x}}_{qp}^T-
          \mathbb{0}\bm{\mathrm{x}}_{qc}^T-
          0
        \bigg)
      =\sum_{q}
        \bigg(\frac{1}{\mathcal{n}}\sum_{n,i}\hat{f}_{niq}\bigg)\cdot1,
  \end{gathered}
\end{gather}
each product's demand gets multiplied with the same contribution margin of one 
and therefore stays unchanged. For competition on revenue, on the other hand, 
only costs must not enter the objective function,
\begin{gather}
  \begin{gathered}
    \label{optrev}
      \sum_{q}
        \bigg(\frac{1}{\mathcal{n}}\sum_{n,i}\hat{f}_{niq}\bigg)
        \bigg(
          \bm{\mathrm{p}}\bm{\mathrm{x}}_{qp}^T-
          \mathbb{0}\bm{\mathrm{x}}_{qc}^T-
          0
        \bigg)
      =\sum_{q}
        \bigg(\frac{1}{\mathcal{n}}\sum_{n,i}\hat{f}_{niq}\bigg)
        \bm{\mathrm{p}}\bm{\mathrm{x}}_{qp}^T,
  \end{gathered}
\end{gather}
which results in each product's demand being multiplied with just price. To 
achieve that in \pkg{cash}, the entries of the price cost matrices have to be
modified accordingly (\Cref{sec:nashgame}), and, for the market share objective, 
the first feature can be treated like any other non-price feature throughout the
framework.

\section{Limitations and future development} 
\label{sec:limdev}
In this paper, we illustrated the implementation and application details of the 
\proglang{R} package \pkg{cash}. To the best of our knowledge, \pkg{cash} is the 
first package to contribute a framework for (end-to-end) simulation of 
competitive reactions based on discrete choice analysis. 

To close off, we would like to address \pkg{cash}'s primary limitations, which 
are congruent with potential avenues for future development. As in our view they 
would be worthwhile, we currently think about extensions to 
\citep{2025_Dressler_et_al}
\setlist{nolistsep}
  \begin{itemize}[noitemsep]
    \item asymmetric competitors (in, e.g., price, cost structure, number of 
          products),
    \item more advanced optimization constraints like reduced manufacturing 
          costs through shared feature levels in a product line 
          \citep{2009_Wang_et_al},
    \item segment heterogeneity \citep[and methods explicitly capturing such 
          preference structures, as finite mixture models, see, e.g.,][] 
          {2002a_Andrews_et_al,2002b_Andrews_et_al,2019_Paetz_et_al,
          2021_Goeken_et_al,2024_Goeken_et_al},
    \item the Stackelberg equilibrium concept \citep[see, e.g.,][]{2010_Steiner},
    \item the integration of a no-choice option, depending upon the definition 
          of its
      \begin{itemize}[noitemsep]
        \item share (assumptions must be made concerning, e.g., the degree of 
              market representation),
        \item attainment (via, e.g., lump-sum after response simulation, 
              calibration of no-choice utility during response simulation) and 
        \item application (only before or also in Nash competition),
        \item [] but note that although it is not a zero-sum game with respect 
              to share anymore when having a no-choice option to gain from or 
              lose to \citep{2012_Chapman_Love}, \cite{2010_Steiner} provides 
              evidence that the no-choice option does not seem to affect the 
              structural properties of the equilibria. Moreover, none of our 
              measures requires the inclusion of a no-choice option for 
              interpretation.
      \end{itemize}
  \end{itemize}
\noindent

Finally, an approach for pruning the vast initial state space (see
\Cref{sec:nashgame}) without loss of equilibrium information would be a 
significant milestone, as the simulation of more complex scenarios might thereby
come within reach.

\section*{Acknowledgments}
The authors thank Philipp Aschersleben for his helpful comments on the 
pre-optimization described in \Cref{sec:nashgame}.

\bibliography{refs}

\begin{thebibliography}{86}
\newcommand{\enquote}[1]{``#1''}
\providecommand{\natexlab}[1]{#1}
\providecommand{\url}[1]{\texttt{#1}}
\providecommand{\urlprefix}{URL }
\expandafter\ifx\csname urlstyle\endcsname\relax
  \providecommand{\doi}[1]{doi:\discretionary{}{}{}#1}\else
  \providecommand{\doi}{doi:\discretionary{}{}{}\begingroup
  \urlstyle{rm}\Url}\fi
\providecommand{\eprint}[2][]{\url{#2}}

\bibitem[{Allenby \emph{et~al.}(2014)Allenby, Brazell, Howell, and
  Rossi}]{2014_Allenby_et_al}
Allenby GM, Brazell JD, Howell JR, Rossi PE (2014).
\newblock \enquote{{Economic Valuation of Product Features}.}
\newblock \emph{Quantitative Marketing and Economics}, \textbf{12}(4),
  421--456.
\newblock \doi{10.1007/s11129-014-9150-x}.

\bibitem[{Allenby and Rossi(1999)}]{1999_Allenby_Rossi}
Allenby GM, Rossi PE (1999).
\newblock \enquote{{Marketing Models of Consumer Heterogeneity}.}
\newblock \emph{Journal of Econometrics}, \textbf{89}(1-2), 57--78.
\newblock \doi{10.1016/S0304-4076(98)00055-4}.

\bibitem[{Andrews \emph{et~al.}(2002{\natexlab{a}})Andrews, Ainslie, and
  Currim}]{2002a_Andrews_et_al}
Andrews RL, Ainslie A, Currim IS (2002{\natexlab{a}}).
\newblock \enquote{{An Empirical Comparison of Logit Choice Models with
  Discrete versus Continuous Representations of Heterogeneity}.}
\newblock \emph{Journal of Marketing Research}, \textbf{39}(4), 479--487.
\newblock \doi{10.1509/jmkr.39.4.479.19124}.

\bibitem[{Andrews \emph{et~al.}(2002{\natexlab{b}})Andrews, Ansari, and
  Currim}]{2002b_Andrews_et_al}
Andrews RL, Ansari A, Currim IS (2002{\natexlab{b}}).
\newblock \enquote{{Hierarchical Bayes versus Finite Mixture Conjoint Analysis
  Models: A Comparison of Fit, Prediction, and Partworth Recovery}.}
\newblock \emph{Journal of Marketing Research}, \textbf{39}(1), 87--98.
\newblock \doi{10.1509/jmkr.39.1.87.18936}.

\bibitem[{Andrews and Currim(2003)}]{2003_Andrews_Currim}
Andrews RL, Currim IS (2003).
\newblock \enquote{{A Comparison of Segment Retention Criteria for Finite
  Mixture Logit Models}.}
\newblock \emph{Journal of Marketing Research}, \textbf{40}(2), 235--243.
\newblock \doi{10.1509/jmkr.40.2.235.19225}.

\bibitem[{Baier and Brusch(2021)}]{2021_Baier_Brusch}
Baier D, Brusch M (eds.) (2021).
\newblock \emph{{Conjointanalyse: Methoden - Anwendungen - Praxisbeispiele}}.
\newblock {Springer-Verlag}.
\newblock \doi{10.1007/978-3-642-00754-5}.

\bibitem[{Baier and Voekler(2024)}]{2024_Baier_Voekler}
Baier D, Voekler S (2024).
\newblock \enquote{{One-Stage Product-Line Design Heuristics: An Empirical
  Comparison}.}
\newblock \emph{OR Spectrum}, \textbf{46}(1), 73--107.
\newblock \doi{10.1007/s00291-023-00716-0}.

\bibitem[{Belloni \emph{et~al.}(2008)Belloni, Freund, Selove, and
  Simester}]{2008_Belloni_et_al}
Belloni A, Freund R, Selove M, Simester D (2008).
\newblock \enquote{{Optimizing Product Line Designs: Efficient Methods and
  Comparisons}.}
\newblock \emph{Management Science}, \textbf{54}(9), 1544--1552.
\newblock \doi{10.1287/mnsc.1080.0864}.

\bibitem[{Bishara and Hittner(2015)}]{2015_Bishara_Hittner}
Bishara AJ, Hittner JB (2015).
\newblock \enquote{{Reducing Bias and Error in the Correlation Coefficient due
  to Nonnormality}.}
\newblock \emph{Educational and Psychological Measurement}, \textbf{75}(5),
  785--804.
\newblock \doi{10.1177/0013164414557639}.

\bibitem[{Brooks and Gelman(1998)}]{1998_Brooks_Gelman}
Brooks SP, Gelman A (1998).
\newblock \enquote{{General Methods for Monitoring Convergence of Iterative
  Simulations}.}
\newblock \emph{Journal of Computational and Graphical Statistics},
  \textbf{7}(4), 434--455.
\newblock \doi{10.1080/10618600.1998.10474787}.

\bibitem[{Brownstone and Train(1999)}]{1999_Brownstone_Train}
Brownstone D, Train K (1999).
\newblock \enquote{{Forecasting New Product Penetration with Flexible
  Substitution Patterns}.}
\newblock \emph{Journal of Econometrics}, \textbf{89}(1-2), 109--129.
\newblock \doi{10.1016/S0304-4076(98)00057-8}.

\bibitem[{Caussade \emph{et~al.}(2005)Caussade, {de Dios Ortúzar}, Rizzi, and
  Hensher}]{2005_Caussade_et_al}
Caussade S, {de Dios Ortúzar} J, Rizzi LI, Hensher DA (2005).
\newblock \enquote{{Assessing the Influence of Design Dimensions on Stated
  Choice Experiment Estimates}.}
\newblock \emph{Transportation Research Part B: Methodological},
  \textbf{39}(7), 621--640.
\newblock \doi{10.1016/j.trb.2004.07.006}.

\bibitem[{Chapman and Love(2012)}]{2012_Chapman_Love}
Chapman CN, Love E (2012).
\newblock \enquote{{Game Theory and Conjoint Analysis: Using Choice Data for
  Strategic Decisions}.}
\newblock In \emph{{Proceedings of the 16th Sawtooth Software Conference}}, pp.
  1--15.

\bibitem[{Choi and DeSarbo(1993)}]{1993_Choi_DeSarbo}
Choi SC, DeSarbo WS (1993).
\newblock \enquote{{Game Theoretic Derivations of Competitive Strategies in
  Conjoint Analysis}.}
\newblock \emph{Marketing Letters}, \textbf{4}(4), 337--348.
\newblock \doi{10.1007/BF00994352}.

\bibitem[{Choi and DeSarbo(1994)}]{1994_Choi_DeSarbo}
Choi SC, DeSarbo WS (1994).
\newblock \enquote{{A Conjoint-Based Product Designing Procedure Incorporating
  Price Competition}.}
\newblock \emph{Journal of Product Innovation Management}, \textbf{11}(5),
  451--459.
\newblock \doi{10.1016/0737-6782(94)90032-9}.

\bibitem[{Choi \emph{et~al.}(1990)Choi, Desarbo, and Harker}]{1990_Choi_et_al}
Choi SC, Desarbo WS, Harker PT (1990).
\newblock \enquote{{Product Positioning under Price Competition}.}
\newblock \emph{Management Science}, \textbf{36}(2), 175--199.
\newblock \doi{10.1287/mnsc.36.2.175}.

\bibitem[{Cook and Nachtsheim(1980)}]{1980_Cook_Nachtsheim}
Cook RD, Nachtsheim CJ (1980).
\newblock \enquote{{A Comparison of Algorithms for Constructing Exact D-Optimal
  Designs}.}
\newblock \emph{Technometrics}, \textbf{22}(3), 315--324.
\newblock \doi{10.1080/00401706.1980.10486162}.

\bibitem[{Cournot(1838)}]{1838_Cournot}
Cournot A (1838).
\newblock \emph{{Recherches sur les Principes Math{\'e}matiques de la
  Th{\'e}orie des Richesses}}.
\newblock {Hachette}.

\bibitem[{Currim \emph{et~al.}(1981)Currim, Weinberg, and
  Wittink}]{1981_Currim_et_al}
Currim IS, Weinberg CB, Wittink DR (1981).
\newblock \enquote{{Design of Subscription Programs for a Performing Arts
  Series}.}
\newblock \emph{Journal of Consumer Research}, \textbf{8}(1), 67--75.
\newblock \doi{10.1086/208842}.

\bibitem[{Dressler \emph{et~al.}(2025)Dressler, Kurz, and
  Steiner}]{2025_Dressler_et_al}
Dressler JHR, Kurz P, Steiner WJ (2025).
\newblock \enquote{{Computing Nash Equilibria for Product Design based on
  Hierarchical Bayesian Mixed Logit Models}.}
\newblock \emph{arXiv preprint (Econometrics)}.
\newblock \doi{10.48550/arXiv.2512.22864}.

\bibitem[{Eddelbuettel \emph{et~al.}(2026)Eddelbuettel, Francois, Allaire,
  Ushey, Kou, Russell, Ucar, Bates, and Chambers}]{2026_Eddelbuettel_et_al}
Eddelbuettel D, Francois R, Allaire JJ, Ushey K, Kou Q, Russell N, Ucar I,
  Bates D, Chambers J (2026).
\newblock \emph{{\pkg{Rcpp}: Seamless \proglang{R} and \proglang{C++}
  Integration}}.
\newblock {\proglang{R} Package}.

\bibitem[{Fedorov(1972)}]{1972_Fedorov}
Fedorov VV (1972).
\newblock \emph{{Theory of Optimal Experiments}}.
\newblock {Academic Press}.

\bibitem[{Fudenberg and Tirole(1991)}]{1991_Fudenberg_Tirole}
Fudenberg D, Tirole J (1991).
\newblock \emph{{Game Theory}}.
\newblock {The MIT Press}.

\bibitem[{Gaujoux(2026)}]{2026_Gaujoux}
Gaujoux R (2026).
\newblock \emph{{\pkg{doRNG}: Generic Reproducible Parallel Backend for
  '\pkg{foreach}' Loops}}.
\newblock {\proglang{R} Package}.

\bibitem[{Gelman \emph{et~al.}(2013)Gelman, Carlin, Stern, Dunson, Vehtari, and
  Rubin}]{2013_Gelman_et_al}
Gelman A, Carlin JB, Stern HS, Dunson DB, Vehtari A, Rubin DB (2013).
\newblock \emph{{Bayesian Data Analysis}}.
\newblock {Chapman \& Hall}.
\newblock \doi{10.1201/b16018}.

\bibitem[{Gelman and Rubin(1992)}]{1992_Gelman_Rubin}
Gelman A, Rubin DB (1992).
\newblock \enquote{{Inference from Iterative Simulation Using Multiple
  Sequences}.}
\newblock \emph{Statistical Science}, \textbf{7}(4), 457--472.
\newblock \doi{10.1214/ss/1177011136}.

\bibitem[{Goeken \emph{et~al.}(2021)Goeken, Kurz, and
  Steiner}]{2021_Goeken_et_al}
Goeken N, Kurz P, Steiner WJ (2021).
\newblock \enquote{{Hierarchical Bayes Conjoint Choice Models – Model
  Framework, Bayesian Inference, Model Selection, and Interpretation of
  Estimation Results}.}
\newblock \emph{Marketing: ZFP – Journal of Research and Management},
  \textbf{43}(3), 49--64.
\newblock \doi{10.15358/0344-1369-2021-3-49}.

\bibitem[{Goeken \emph{et~al.}(2024)Goeken, Kurz, and
  Steiner}]{2024_Goeken_et_al}
Goeken N, Kurz P, Steiner WJ (2024).
\newblock \enquote{{Multimodal Preference Heterogeneity in Choice-Based
  Conjoint Analysis: A Simulation Study}.}
\newblock \emph{Journal of Business Economics}, \textbf{94}(1), 137--185.
\newblock \doi{10.1007/s11573-023-01156-6}.

\bibitem[{Green and Krieger(1997)}]{1997_Green_Krieger}
Green PE, Krieger AM (1997).
\newblock \emph{{Using Conjoint Analysis to View Competitive Interaction
  through the Customer’s Eyes}}, p. 343–367.
\newblock {John Wiley \& Sons}.

\bibitem[{Green and Rao(1971)}]{1971_Green_Rao}
Green PE, Rao VR (1971).
\newblock \enquote{{Conjoint Measurement for Quantifying Judgmental Data}.}
\newblock \emph{Journal of Marketing Research}, \textbf{8}(3), 355--363.
\newblock \doi{10.1177/002224377100800312}.

\bibitem[{Green and Srinivasan(1990)}]{1990_Green_Srinivasan}
Green PE, Srinivasan V (1990).
\newblock \enquote{{Conjoint Analysis in Marketing: New Developments with
  Implications for Research and Practice}.}
\newblock \emph{Journal of Marketing}, \textbf{54}(4), 3--19.
\newblock \doi{10.1177/002224299005400402}.

\bibitem[{Gr{\"o}mping(2018)}]{2018_Groemping}
Gr{\"o}mping U (2018).
\newblock \enquote{{\proglang{R} Package \pkg{DoE.base} for Factorial
  Experiments}.}
\newblock \emph{Journal of Statistical Software}, \textbf{85}(5), 1--41.
\newblock \doi{10.18637/jss.v085.i05}.

\bibitem[{Gustafsson \emph{et~al.}(2007)Gustafsson, Herrmann, and
  Huber}]{2007_Gustafsson_et_al}
Gustafsson A, Herrmann A, Huber F (eds.) (2007).
\newblock \emph{{Conjoint Measurement: Methods and Applications}}.
\newblock {Springer-Verlag}.
\newblock \doi{10.1007/978-3-540-71404-0}.

\bibitem[{Gutsche(1995)}]{1995_Gutsche}
Gutsche J (1995).
\newblock \emph{{Produktpräferenzanalyse: Ein Modelltheoretisches und
  Methodisches Konzept zur Marktsimulation mittels
  Präferenzerfassungsmodellen}}.
\newblock {Duncker \& Humblot}.

\bibitem[{Haaijer and Wedel(2007)}]{2007_Haaijer_Wedel}
Haaijer R, Wedel M (2007).
\newblock \emph{{Conjoint Choice Experiments: General Characteristics and
  Alternative Model Specifications}}, pp. 199--229.
\newblock {Springer-Verlag}.
\newblock \doi{10.1007/978-3-540-71404-0}.

\bibitem[{Hein \emph{et~al.}(2022)Hein, Goeken, Kurz, and
  Steiner}]{2022_Hein_et_al}
Hein M, Goeken N, Kurz P, Steiner WJ (2022).
\newblock \enquote{{Using Hierarchical Bayes Draws for Improving Shares of
  Choice Predictions in Conjoint Simulations: A Study Based on Conjoint Choice
  Data}.}
\newblock \emph{European Journal of Operational Research}, \textbf{297}(2),
  630--651.
\newblock \doi{10.1016/j.ejor.2021.05.056}.

\bibitem[{Hein \emph{et~al.}(2019)Hein, Kurz, and Steiner}]{2019_Hein_et_al}
Hein M, Kurz P, Steiner WJ (2019).
\newblock \enquote{{On the Effect of HB Covariance Matrix Prior Settings: A
  Simulation Study}.}
\newblock \emph{Journal of Choice Modelling}, \textbf{31}, 51--72.
\newblock \doi{10.1016/j.jocm.2019.02.001}.

\bibitem[{Hein \emph{et~al.}(2020)Hein, Kurz, and Steiner}]{2020_Hein_et_al}
Hein M, Kurz P, Steiner WJ (2020).
\newblock \enquote{{Analyzing the Capabilities of the HB Logit Model for
  Choice-Based Conjoint Analysis: A Simulation Study}.}
\newblock \emph{Journal of Business Economics}, \textbf{90}(1), 1--36.
\newblock \doi{10.1007/s11573-019-00927-4}.

\bibitem[{Hensher \emph{et~al.}(2015)Hensher, Rose, and
  Greene}]{2015_Hensher_et_al}
Hensher DA, Rose JM, Greene WH (2015).
\newblock \emph{{Applied Choice Analysis}}.
\newblock {Cambridge University Press}.
\newblock \doi{10.1017/CBO9781316136232}.

\bibitem[{Hensher \emph{et~al.}(2001)Hensher, Stopher, and
  Louviere}]{2001_Hensher_et_al}
Hensher DA, Stopher PR, Louviere JJ (2001).
\newblock \enquote{{An Exploratory Analysis of the Effect of Numbers of Choice
  Sets in Designed Choice Experiments: An Airline Choice Application}.}
\newblock \emph{Journal of Air Transport Management}, \textbf{7}(6), 373--379.
\newblock \doi{10.1016/S0969-6997(01)00031-X}.

\bibitem[{Hess and Daly(2024)}]{2024_Hess_Daly}
Hess S, Daly A (eds.) (2024).
\newblock \emph{{Handbook of Choice Modelling}}.
\newblock {Edward Elgar Publishing}.
\newblock \doi{10.4337/9781800375635}.

\bibitem[{Hoogerbrugge and {van der Wagt}(2006)}]{2006_Hoogerbrugge_vanderWagt}
Hoogerbrugge M, {van der Wagt} K (2006).
\newblock \enquote{{How Many Choice Tasks Should We Ask?}}
\newblock In \emph{{Proceedings of the 12th Sawtooth Software Conference}}, pp.
  97--110.

\bibitem[{Huber and Zwerina(1996)}]{1996_Huber_Zwerina}
Huber J, Zwerina K (1996).
\newblock \enquote{{The Importance of Utility Balance in Efficient Choice
  Designs}.}
\newblock \emph{Journal of Marketing Research}, \textbf{33}(3), 307--317.
\newblock \doi{10.1177/002224379603300305}.

\bibitem[{Johnson and Orme(1996)}]{1996_Johnson_Orme}
Johnson RM, Orme BK (1996).
\newblock \enquote{{How Many Questions Should You Ask in Choice-Based Conjoint
  Studies?}}
\newblock \emph{Technical report}, {Sawtooth Software, Inc.}, {Sequim, WA,
  USA}.

\bibitem[{Kohli and Sukumar(1990)}]{1990_Kohli_Sukumar}
Kohli R, Sukumar R (1990).
\newblock \enquote{{Heuristics for Product-Line Design Using Conjoint
  Analysis}.}
\newblock \emph{Management Science}, \textbf{36}(12), 1464--1478.
\newblock \doi{10.1287/mnsc.36.12.1464}.

\bibitem[{Kuhfeld(2010)}]{2010_Kuhfeld}
Kuhfeld WF (2010).
\newblock \emph{{The Macros}}, pp. 803--1212.
\newblock {\proglang{SAS} Institute Inc.}, {Cary, NC, USA}.

\bibitem[{Kurz and Binner(2012)}]{2012_Kurz_Binner}
Kurz P, Binner S (2012).
\newblock \enquote{{"The Individual Choice Task Threshold". Need for Variable
  Number of Choice Tasks}.}
\newblock In \emph{{Proceedings of the 16th Sawtooth Software Conference}}, pp.
  111--128.

\bibitem[{Kuzmanovic and Martic(2012)}]{2012_Kuzmanovic_Martic}
Kuzmanovic M, Martic M (2012).
\newblock \enquote{{An Approach to Competitive Product Line Design Using
  Conjoint Data}.}
\newblock \emph{Expert Systems with Applications}, \textbf{39}(8), 7262--7269.
\newblock \doi{10.1016/j.eswa.2012.01.097}.

\bibitem[{Kuzmanovic \emph{et~al.}(2019)Kuzmanovic, Martic, and
  Vujosevic}]{2019_Kuzmanovic_et_al}
Kuzmanovic M, Martic M, Vujosevic M (2019).
\newblock \enquote{{Designing a Profit-Maximizing Product Line for
  Heterogeneous Market}.}
\newblock \emph{Technical Gazette}, \textbf{26}(6), 1562--1569.
\newblock \doi{10.17559/TV-20180811192832}.

\bibitem[{Louviere \emph{et~al.}(2013)Louviere, Carson, Burgess, Street, and
  Marley}]{2013_Louviere_et_al}
Louviere JJ, Carson RT, Burgess L, Street D, Marley AAJ (2013).
\newblock \enquote{{Sequential Preference Questions Factors Influencing
  Completion Rates and Response Times Using an Online Panel}.}
\newblock \emph{Journal of Choice Modelling}, \textbf{8}, 19--31.
\newblock \doi{10.1016/j.jocm.2013.04.009}.

\bibitem[{Louviere \emph{et~al.}(2000)Louviere, Hensher, and
  Swait}]{2000_Louviere_et_al}
Louviere JJ, Hensher DA, Swait JD (2000).
\newblock \emph{{Stated Choice Methods: Analysis and Applications}}.
\newblock {Cambridge University Press}.
\newblock \doi{10.1017/CBO9780511753831}.

\bibitem[{Louviere and Woodworth(1983)}]{1983_Louviere_Woodworth}
Louviere JJ, Woodworth G (1983).
\newblock \enquote{{Design and Analysis of Simulated Consumer Choice or
  Allocation Experiments: An Approach Based on Aggregate Data}.}
\newblock \emph{Journal of Marketing Research}, \textbf{20}(4), 350--367.
\newblock \doi{10.1177/002224378302000403}.

\bibitem[{McFadden(1974)}]{1974_McFadden}
McFadden D (1974).
\newblock \emph{{Conditional Logit Analysis of Qualitative Choice Behavior}},
  p. 105–142.
\newblock {Academic Press}.

\bibitem[{{Microsoft Corporation} and
  Weston(2022{\natexlab{a}})}]{2022a_Microsoft_Weston}
{Microsoft Corporation}, Weston S (2022{\natexlab{a}}).
\newblock \emph{{\pkg{doParallel}: Foreach Parallel Adaptor for the
  '\pkg{parallel}' Package}}.
\newblock {\proglang{R} Package}.

\bibitem[{{Microsoft Corporation} and
  Weston(2022{\natexlab{b}})}]{2022b_Microsoft_Weston}
{Microsoft Corporation}, Weston S (2022{\natexlab{b}}).
\newblock \emph{{\pkg{foreach}: Provides Foreach Looping Construct}}.
\newblock {\proglang{R} Package}.

\bibitem[{Nash(1951)}]{1951_Nash}
Nash J (1951).
\newblock \enquote{{Non-Cooperative Games}.}
\newblock \emph{Annals of Mathematics}, \textbf{54}(2), 286--295.
\newblock \doi{10.2307/1969529}.

\bibitem[{Orme and Baker(2000)}]{2000_Orme_Baker}
Orme B, Baker G (2000).
\newblock \enquote{{Comparing Hierarchical Bayes Draws and Randomized First
  Choice for Conjoint Simulations}.}
\newblock In \emph{{Proceedings of the 8th Sawtooth Software Conference}}, pp.
  239--254.

\bibitem[{Paetz \emph{et~al.}(2019)Paetz, Hein, Kurz, and
  Steiner}]{2019_Paetz_et_al}
Paetz F, Hein M, Kurz P, Steiner WJ (2019).
\newblock \enquote{{Latent Class Conjoint Choice Models: A Guide for Model
  Selection, Estimation, Validation, and Interpretation of Results}.}
\newblock \emph{Marketing: ZFP – Journal of Research and Management},
  \textbf{41}(4), 3--20.
\newblock \doi{10.15358/0344-1369-2019-4-3}.

\bibitem[{Pinnell and Englert(1997)}]{1997_Pinnell_Englert}
Pinnell J, Englert S (1997).
\newblock \enquote{{The Number of Choice Alternatives in Discrete Choice
  Modeling}.}
\newblock In \emph{{Proceedings of the 6th Sawtooth Software Conference}}, pp.
  121--154.

\bibitem[{Plummer \emph{et~al.}(2006)Plummer, Best, Cowles, and
  Vines}]{2006_Plummer_et_al}
Plummer M, Best N, Cowles K, Vines K (2006).
\newblock \enquote{{\pkg{CODA}: Convergence Diagnosis and Output Analysis for
  MCMC}.}
\newblock \emph{\proglang{R} News}, \textbf{6}(1), 7--11.

\bibitem[{Rao(2014)}]{2014_Rao}
Rao VR (2014).
\newblock \emph{{Applied Conjoint Analysis}}.
\newblock {Springer-Verlag}.
\newblock \doi{10.1007/978-3-540-87753-0}.

\bibitem[{{\proglang{R} Core Team}(2026)}]{2026_RCoreTeam}
{\proglang{R} Core Team} (2026).
\newblock \emph{{\proglang{R}: A Language and Environment for Statistical
  Computing}}.
\newblock {\proglang{R} Foundation for Statistical Computing}.

\bibitem[{Rose and Bliemer(2009)}]{2009_Rose_Bliemer}
Rose JM, Bliemer MCJ (2009).
\newblock \enquote{{Constructing Efficient Stated Choice Experimental
  Designs}.}
\newblock \emph{Transport Reviews}, \textbf{29}(5), 587--617.
\newblock \doi{10.1080/01441640902827623}.

\bibitem[{Rossi(2025)}]{2025_Rossi}
Rossi PE (2025).
\newblock \emph{{\pkg{bayesm}: Bayesian Inference for
  Marketing/Micro-Econometrics}}.
\newblock {\proglang{R} Package}.

\bibitem[{Rossi \emph{et~al.}(2005)Rossi, Allenby, and
  McCulloch}]{2005_Rossi_et_al}
Rossi PE, Allenby GM, McCulloch R (2005).
\newblock \emph{{Bayesian Statistics and Marketing}}.
\newblock {John Wiley \& Sons}.
\newblock \doi{10.1002/9781394219148}.

\bibitem[{Shocker and Srinivasan(1979)}]{1979_Shocker_Srinivasan}
Shocker AD, Srinivasan V (1979).
\newblock \enquote{{Multiattribute Approaches for Product Concept Evaluation
  and Generation: A Critical Review}.}
\newblock \emph{Journal of Marketing Research}, \textbf{16}(2), 159--180.
\newblock \doi{10.1177/002224377901600202}.

\bibitem[{Silverman(1986)}]{1986_Silverman}
Silverman BW (1986).
\newblock \emph{{Density Estimation for Statistics and Data Analysis}}.
\newblock {Chapman \& Hall}.
\newblock \doi{10.1002/bimj.4710300745}.

\bibitem[{Srinivasan(1975)}]{1975_Srinivasan}
Srinivasan V (1975).
\newblock \enquote{{Linear Programming Computational Procedures for Ordinal
  Regression}.}
\newblock \emph{Journal of the Association for Computing Machinery},
  \textbf{23}(3), 475--487.
\newblock \doi{10.1145/321958.321969}.

\bibitem[{Steiner(2010)}]{2010_Steiner}
Steiner WJ (2010).
\newblock \enquote{{A Stackelberg-Nash Model for New Product Design}.}
\newblock \emph{OR Spectrum}, \textbf{32}(1), 21--48.
\newblock \doi{10.1007/s00291-008-0137-4}.

\bibitem[{Steiner and Hruschka(2000)}]{2000_Steiner_Hruschka}
Steiner WJ, Hruschka H (2000).
\newblock \enquote{{Conjoint-Based Product (Line) Design Considering
  Competitive Reactions}.}
\newblock \emph{OR Spectrum}, \textbf{22}(1), 71--95.
\newblock \doi{10.1007/s002910050006}.

\bibitem[{Stephenson(2002)}]{2002_Stephenson}
Stephenson AG (2002).
\newblock \enquote{{\pkg{evd}: Extreme Value Distributions}.}
\newblock \emph{\proglang{R} News}, \textbf{2}(2), 31--32.

\bibitem[{Street and Viney(2019)}]{2019_Street_Viney}
Street DJ, Viney R (2019).
\newblock \emph{{Design of Discrete Choice Experiments}}.
\newblock {Oxford University Press}.
\newblock \doi{10.1093/acrefore/9780190625979.013.91}.

\bibitem[{Traets \emph{et~al.}(2020)Traets, Sanchez, and
  Vandebroek}]{2020_Traets_et_al}
Traets F, Sanchez DG, Vandebroek M (2020).
\newblock \enquote{{Generating Optimal Designs for Discrete Choice Experiments
  in \proglang{R}: The \pkg{idefix} Package}.}
\newblock \emph{Journal of Statistical Software}, \textbf{96}(3), 1--41.
\newblock \doi{10.18637/jss.v096.i03}.

\bibitem[{Train(2009)}]{2009_Train}
Train KE (2009).
\newblock \emph{{Discrete Choice Methods with Simulation}}.
\newblock {Cambridge University Press}.
\newblock \doi{10.1017/CBO9780511805271}.

\bibitem[{{von Stackelberg}(1934)}]{1934_vonStackelberg}
{von Stackelberg} H (1934).
\newblock \emph{{Marktform und Gleichgewicht}}.
\newblock {Springer-Verlag}.
\newblock \doi{10.1007/978-3-642-12586-7}.

\bibitem[{Vriens \emph{et~al.}(1996)Vriens, Wedel, and
  Wilms}]{1996_Vriens_et_al}
Vriens M, Wedel M, Wilms T (1996).
\newblock \enquote{{Metric Conjoint Segmentation Methods: A Monte Carlo
  Comparison}.}
\newblock \emph{Journal of Marketing Research}, \textbf{33}(1), 73--85.
\newblock \doi{10.1177/002224379603300107}.

\bibitem[{Walker \emph{et~al.}(2018)Walker, Wang, Thorhauge, and
  Ben-Akiva}]{2018_Walker_et_al}
Walker JL, Wang Y, Thorhauge M, Ben-Akiva M (2018).
\newblock \enquote{{D-Efficient or Deficient? A Robustness Analysis of Stated
  Choice Experimental Designs}.}
\newblock \emph{Theory and Decision}, \textbf{84}(2), 215--238.
\newblock \doi{10.1007/s11238-017-9647-3}.

\bibitem[{Wang \emph{et~al.}(2009)Wang, Camm, and Curry}]{2009_Wang_et_al}
Wang XJ, Camm JD, Curry DJ (2009).
\newblock \enquote{{A Branch-and-Price Approach to the Share-of-Choice Product
  Line Design Problem}.}
\newblock \emph{Management Science}, \textbf{55}(10), 1718--1728.
\newblock \doi{10.1287/mnsc.1090.1058}.

\bibitem[{Wedel and Steenkamp(1989)}]{1989_Wedel_Steenkamp}
Wedel M, Steenkamp JBEM (1989).
\newblock \enquote{{A Fuzzy Clusterwise Regression Approach to Benefit
  Segmentation}.}
\newblock \emph{International Journal of Research in Marketing}, \textbf{6}(4),
  241--258.
\newblock \doi{10.1016/0167-8116(89)90052-9}.

\bibitem[{Wickham(2016)}]{2016_Wickham}
Wickham H (2016).
\newblock \emph{{\pkg{ggplot2}: Elegant Graphics for Data Analysis}}.
\newblock {Springer-Verlag}.
\newblock \doi{10.1007/978-3-319-24277-4}.

\bibitem[{Wirth(2010{\natexlab{a}})}]{2010a_Wirth}
Wirth R (2010{\natexlab{a}}).
\newblock \emph{{Best-Worst Choice-Based Conjoint-Analyse. Eine Neue Variante
  der Wahlbasierten Conjoint-Analyse}}.
\newblock {Tectum}.

\bibitem[{Wirth(2010{\natexlab{b}})}]{2010b_Wirth}
Wirth R (2010{\natexlab{b}}).
\newblock \enquote{{HB-CBC, HB-Best-Worst-CBC or No HB At All?}}
\newblock In \emph{{Proceedings of the 15th Sawtooth Software Conference}}, pp.
  321--356.

\bibitem[{Wittink and Cattin(1981)}]{1981_Wittink_Cattin}
Wittink DR, Cattin P (1981).
\newblock \enquote{{Alternative Estimation Methods for Conjoint Analysis: A
  Monté Carlo Study}.}
\newblock \emph{Journal of Marketing Research}, \textbf{18}(1), 101--106.
\newblock \doi{10.1177/002224378101800112}.

\bibitem[{Wittink \emph{et~al.}(1982)Wittink, Krishnamurthi, and
  Nutter}]{1982_Wittink_et_al}
Wittink DR, Krishnamurthi L, Nutter JB (1982).
\newblock \enquote{{Comparing Derived Importance Weights Across Attributes}.}
\newblock \emph{Journal of Consumer Research}, \textbf{8}(4), 471--474.
\newblock \doi{10.1086/208890}.

\bibitem[{Wittink \emph{et~al.}(1990)Wittink, Krishnamurthi, and
  Reibstein}]{1990_Wittink_et_al}
Wittink DR, Krishnamurthi L, Reibstein DJ (1990).
\newblock \enquote{{The Effect of Differences in the Number of Attribute Levels
  on Conjoint Results}.}
\newblock \emph{Marketing Letters}, \textbf{1}(2), 113--123.
\newblock \doi{10.1007/BF00435295}.

\bibitem[{Zwerina \emph{et~al.}(2010)Zwerina, Huber, and
  Kuhfeld}]{2010_Zwerina_et_al}
Zwerina K, Huber J, Kuhfeld WF (2010).
\newblock \emph{{A General Method for Constructing Efficient Choice Designs}},
  pp. 265--284.
\newblock {\proglang{SAS} Institute Inc.}, {Cary, NC, USA}.

\end{thebibliography}

\end{document}